\documentclass[12pt]{article}

\usepackage[left=2cm,right=2cm,top=1.5cm,bottom=2cm]{geometry}

\usepackage{graphicx}
\usepackage[dvipsnames]{xcolor}
\usepackage{colortbl}
\usepackage{enumitem}
\usepackage[none]{hyphenat}
\usepackage{placeins}
\usepackage{authblk}
\usepackage{dsfont}
\usepackage{longtable}
\usepackage{tabularx}
\usepackage{float}
\usepackage{multirow}
\usepackage{booktabs}
\usepackage{makecell}
\usepackage{bm}
\usepackage{tikz}
\usetikzlibrary{decorations.pathreplacing}

\usepackage[fleqn]{amsmath}
\usepackage{amssymb}
\usepackage{nccmath}

\usepackage{caption}
\usepackage{subcaption}
\usepackage{soul}

\usepackage[linesnumbered,ruled,vlined]{algorithm2e}

\usepackage[authoryear]{natbib}
\usepackage[hidelinks]{hyperref}
\hypersetup{
  colorlinks=true,
  linkcolor=blue,
  filecolor=black,
  citecolor=blue
}

\title{Dynamic Mortality Forecasting via Mixed-Frequency State-Space Models}
\author[1]{Runze Li\footnote{Corresponding Author. runze2@student.unimelb.edu.au}}
\author[1]{Rui Zhou\footnote{rui.zhou@unimelb.edu.au}}
\author[1]{David Pitt\footnote{david.pitt@unimelb.edu.au}}
\affil[1]{Department of Economics, University of Melbourne}
\date{}
\begin{document}

\maketitle
\section*{Abstract}

High-frequency death counts are now widely available and contain timely information about intra-year mortality dynamics, but most stochastic mortality models are still estimated on annual data and therefore update only when annual totals are released. We propose a mixed-frequency state-space (MF--SS) extension of the Lee--Carter framework that jointly uses annual mortality rates and monthly death counts. The two series are linked through a shared latent monthly mortality factor, with the annual period factor defined as the intra-year average of the monthly factors. The latent monthly factor follows a seasonal ARIMA process, and parameters are estimated by maximum likelihood using an EM algorithm with Kalman filtering and smoothing. This setup enables real-time intra-year updates of the latent state and forecasts as new monthly observations arrive without re-estimating model parameters.

Using U.S.\ data for ages 20--90 over 1999--2019, we evaluate intra-year annual nowcasts and one- to five-year-ahead forecasts. The MF--SS model produces both a direct annual forecast and an annual forecast implied by aggregating monthly projections. In our application, the aggregated monthly forecast is typically more accurate. Incorporating monthly information substantially improves intra-year annual nowcasts, especially after the first few months of the year. As a benchmark, we also fit separate annual and monthly Lee--Carter models and combine their forecasts using temporal reconciliation. Reconciliation improves these independent forecasts but adds little to MF--SS forecasts, consistent with MF--SS pooling information across frequencies during estimation. The MF--SS aggregated monthly forecasts generally outperform both unreconciled and temporally reconciled Lee--Carter forecasts and produce more cautious predictive intervals than the reconciled Lee--Carter approach.

\bigskip

\noindent
\textbf{Keywords:} mortality forecasting; mixed-frequency data; state-space models; Kalman filter; EM algorithm; seasonal ARIMA; temporal forecast reconciliation.

\section{Introduction}
Accurate mortality forecasts are a core input to life insurance pricing, pension valuation, and public health planning. The seminal Lee--Carter model \citep{leeandcarter1992} and the Cairns--Blake--Dowd (CBD) model \citep{Cairns2006}, along with their various extensions \citep[e.g.][]{RENSHAW2006,currie2006,Cairns2009}, are typically estimated using annual mortality data. Annual data provide a stable signal of long-run mortality change, but they are released with delay and therefore update slowly when mortality conditions shift. At the same time, many statistical agencies now publish high-frequency death counts (weekly or monthly), which are timely and reveal strong intra-year patterns such as seasonality. Because high-frequency counts are less aggregated, they are typically more variable. Early in the calendar year, annual nowcasts based on monthly data can change more from month to month because most of the year remains unobserved and must be forecast, and each new month updates those remaining-month projections \citep{rli2026}. 

This paper addresses a practical forecasting question: \emph{how can we use both stable annual mortality rates and timely monthly death counts within a single framework that delivers accurate forecasts and supports real-time intra-year updating?}

Our main contribution is a mixed-frequency state-space (MF--SS) mortality model that extends the Lee--Carter structure to jointly use annual mortality rates and monthly death counts. The model includes two Lee--Carter-like observation equations---one for annual log mortality rates and one for a scaled monthly log series constructed from monthly deaths---linked through a shared latent monthly mortality factor. The annual period factor is defined as the intra-year average of the monthly factors, providing a structural cross-frequency link between annual and monthly dynamics. The latent monthly factor follows a seasonal ARIMA (SARIMA) transition equation, capturing long-run drift, short-run persistence, and seasonal dynamics. Because the system is linear and Gaussian, we estimate parameters by maximum likelihood using an EM algorithm with Kalman filtering and smoothing \citep{kalmanRE1960,koopman2012}. Once parameters are estimated, new monthly observations can be incorporated through Kalman filtering, updating the latent state and forecasts within the year without re-estimating the full model.

The MF--SS approach builds on a well-established state-space literature for mortality modelling, where Lee--Carter-type specifications are cast in latent-state form to enable likelihood-based inference and state filtering/smoothing. \citet{pedroza2006} formulates a Bayesian Lee--Carter model and illustrates how the state-space structure facilitates uncertainty quantification and the handling of missing data. \citet{fung2015} develop a Bayesian state-space representation of Lee--Carter and discuss implications for long-horizon quantities such as annuity values. More general state-space formulations and identification issues are studied in \citet{Fung2017}, while \citet{Fung2019} incorporate cohort effects within a Bayesian state-space framework. \citet{LIU2016301} and \citet{LI202396} develop longevity risk hedging strategies based on general state-space representations of stochastic mortality models. Beyond discrete-time specifications, \citet{andersson2021} propose a Lexis-based approach using continuous-time mortality dynamics. These studies demonstrate the flexibility of state-space methods for mortality forecasting, but they are predominantly single-frequency (annual) and do not directly address joint modelling of annual outcomes together with high-frequency death counts. The closest related work is \citet[Chapter~4]{shi2024thesis}, which develops
a Bayesian mixed-frequency FAVAR model linking annual mortality rates with
quarterly macroeconomic indicators to infer quarterly mortality and study
economic shocks. As in our model, a latent higher-frequency mortality factor
is tied to annual observations through an aggregation constraint
\citep{murasawa2003}. In contrast, we use timely but noisy monthly death
counts alongside annual data to improve mortality nowcasting and
forecasting. Our joint Lee--Carter-type state-space model, estimated by EM
with Kalman filtering and smoothing, updates forecasts as new monthly data
arrive without re-estimation and produces both annual and monthly forecasts.

At the same time, an emerging literature shows that high-frequency mortality data are informative about short-run dynamics, seasonality, and mortality shocks. \citet{robben2025a} relate weekly mortality deviations from seasonal baselines to environmental conditions such as weather and air pollution, while \citet{robben2025b} develop a regime-switching framework that separates baseline mortality from heat-wave and respiratory-disease shock regimes using weekly data and external indicators. In the context of COVID-19, \citet{vanberkum2025} extend the Li--Lee framework with a pandemic component calibrated using weekly Short-Term Mortality Fluctuations data. These contributions highlight the value of temporal granularity, but they typically either focus on the high-frequency series itself or combine annual and high-frequency information in multi-stage ways rather than estimating both frequencies jointly.

In related work, \citet{rli2026} propose a Mixed Data Sampling (MIDAS) approach that uses monthly death counts as high-frequency regressors to forecast and update annual mortality rates as new months of data arrive. MIDAS provides a convenient reduced-form mechanism for annual nowcasting, but it targets annual outcomes only and does not produce corresponding monthly forecasts. Moreover, because the MIDAS mapping depends on the forecast horizon and the intra-year information set, it is typically implemented via a collection of month- and horizon-specific regressions. In contrast, the MF--SS framework specifies a single joint model for annual and monthly observations linked by a shared latent monthly factor. Once parameters are estimated, intra-year updates are carried out by filtering rather than repeated re-estimation.

Mixed-frequency state-space models are widely used in macroeconomic nowcasting to combine low- and high-frequency measurements of a common latent process. \citet{murasawa2003} link monthly indicators to quarterly GDP within a state-space dynamic factor model, while \citet{mm2016} allow for time-varying volatility in the latent dynamics and \citet{CHERNIS2020851} incorporate multiple observation equations to accommodate three sampling frequencies. These developments motivate our use of mixed-frequency state-space methods for mortality, where annual and monthly observations provide different-frequency measurements of a shared latent mortality component.

To benchmark the value of joint modelling, we also consider an alternative: fitting separate Lee--Carter models to the annual and monthly series and then combining the resulting forecasts using temporal forecast reconciliation. We adopt the optimal-combination framework of \citet{hyndman2011} in its temporal-hierarchy form as proposed by \citet{ATHANASOPOULOS201760}, which combines independently generated annual and monthly base forecasts subject. This benchmark is simple to implement, but unlike MF--SS it does not pool information across frequencies during estimation.

Temporal reconciliation is closely related in spirit to a growing mortality-forecasting literature that enforces aggregation consistency across grouped (cross-sectional) structures, such as sex, region, or subpopulations. In particular, \citet{shanghyn2017} introduce grouped functional time-series forecasting for age-specific mortality and show that reconciling forecasts across grouped structures can improve accuracy while maintaining aggregation consistency; see also \citet{shanghaberman2017} for multivariate extensions and \citet{shang2020} for a comparative evaluation of reconciliation and forecast-combination strategies in grouped mortality settings.  \citet{LI2019122} incorporate the disaggregate cause-specific mortality data and the aggregate total-level data via reconciliation and achieve better forecasting results at both levels.  While these studies focus on cross-sectional aggregation structures, our application uses the same reconciliation logic in the temporal dimension (annual versus monthly) and allows a direct comparison between ex-post reconciliation and joint likelihood-based pooling via the MF--SS framework.

Empirically, we apply both approaches to U.S.\ mortality data for ages 20--90 over 1999--2019, the period where annual mortality rates and monthly death counts overlap. We evaluate annual mortality forecasting performance using intra-year nowcasts and multi-year-ahead forecasts. Our results show that (i) incorporating monthly information materially improves intra-year annual nowcasts, especially after the earliest months of the year; (ii) the mixed-frequency state-space model improves annual forecast accuracy relative to single-frequency Lee--Carter benchmarks; and (iii) temporal reconciliation substantially improves forecasts from separately estimated annual/monthly Lee--Carter models, but adds little to the state-space forecasts, consistent with the MF--SS model already pooling information across frequencies.

In summary, the paper makes four contributions:
\begin{enumerate}
  \item We propose a mixed-frequency Lee--Carter state-space model that jointly uses annual mortality rates and monthly death counts, linking annual and monthly dynamics through the latent period factor.
  \item We develop a practical estimation and intra-year updating strategy based on EM and Kalman filtering/smoothing that updates forecasts as new monthly data arrive without re-estimating parameters each month.
  \item We provide a transparent benchmark based on independently estimated annual and monthly Lee--Carter models combined via temporal forecast reconciliation.
  \item Using U.S.\ data, we quantify the gains from mixed-frequency modelling for annual mortality nowcasting and multi-year forecasting, and clarify when reconciliation helps and when it does not.
\end{enumerate}

The remainder of the paper is organised as follows. Section~\ref{sec:data} describes the annual and monthly data and the construction of a comparable monthly series. Section~\ref{sec:mfss_model} presents the mixed-frequency state-space model, its state-space representation, and the EM/Kalman estimation and forecasting procedures. Section~\ref{sec:ind_LC} introduces the independent annual and monthly Lee--Carter benchmarks and the intra-year updating strategy for monthly factors. Section~\ref{sec:results} reports parameter estimates and evaluates forecast accuracy across horizons and intra-year information sets. Section~\ref{sec:reconciliation} discusses temporal forecast reconciliation and compares reconciled and unreconciled forecasts. Section~\ref{sec:conclude} concludes and outlines extensions.

\section{Data}
\label{sec:data}

The analysis uses U.S. mortality data, with annual mortality rates and initial population exposures obtained from the Human Mortality Database (HMD) \citep{HMD2025} and monthly death counts drawn from the Centers for Disease Control and Prevention (CDC) \citep{CDC2025}. HMD provides annual mortality rates at single-year ages from 1933 onwards, while the CDC supplies monthly death counts at single-year ages from January 1999. Our analysis focuses on the period January 1999 to December 2019 for ages 20 to 90, over which both monthly death counts and annual mortality rates are available. 
The pandemic period is excluded to avoid distortions from mortality shocks that are not representative of the underlying dynamics in pre-pandemic period. 

\begin{figure}[H]
    \centering
    \caption{Annual mortality rates and scaled monthly death rates ($\times 12$)  on a log scale, 1999--2019, averaged over ages 20-90. }
    \includegraphics[width=0.8\textwidth]{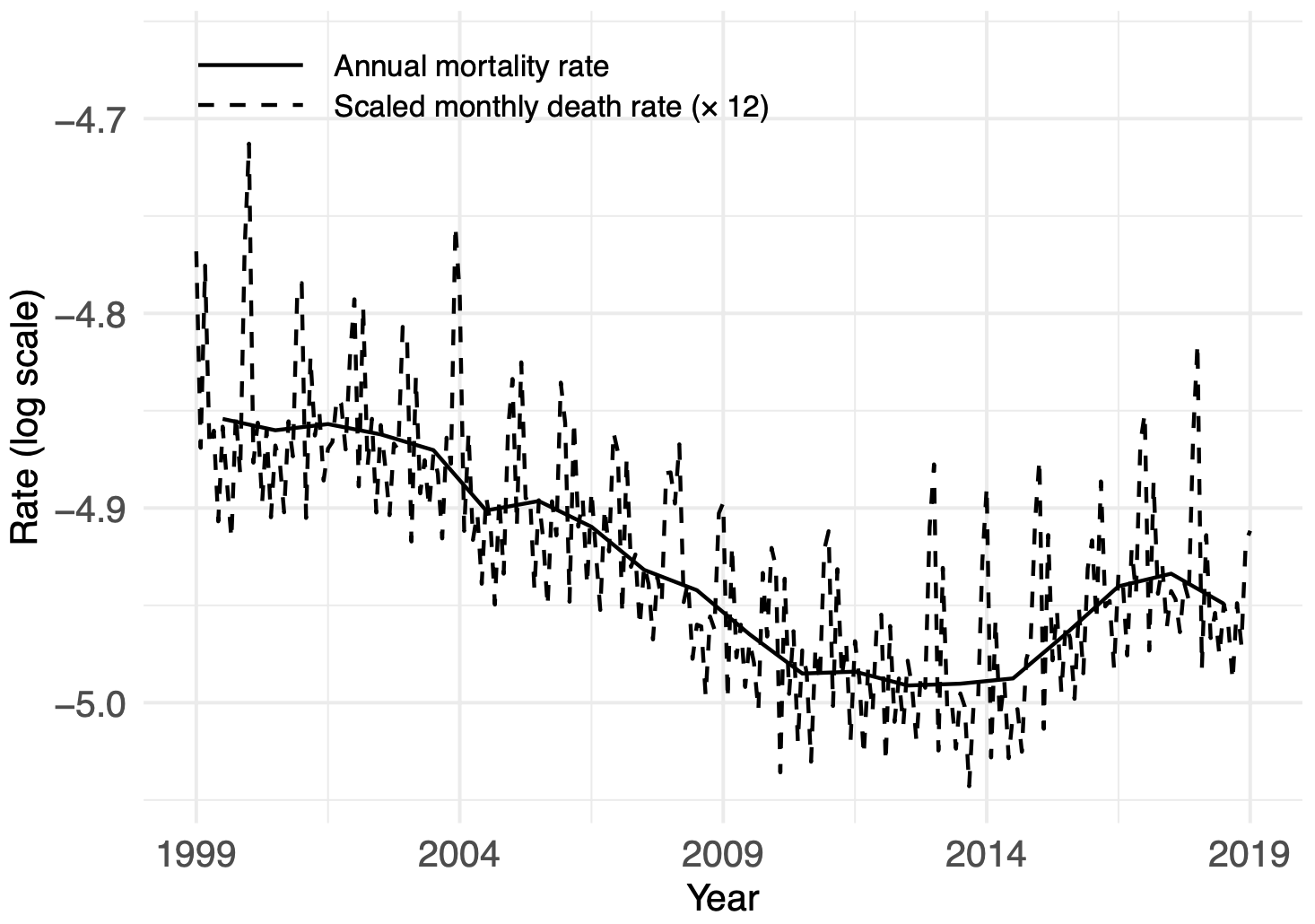}
  
    \label{fig:data_plot}
\end{figure}

Figure \ref{fig:data_plot} displays the annual mortality rates and the scaled monthly death rates on a log scale, each averaged across ages. For age $x$ in year $t$, we construct the scaled monthly death rate as monthly deaths divided by the initial population at age $x$ at the beginning of year $t$. For presentation, we plot 12 times this quantity so that it is on a comparable scale to the annual mortality rates. The figure shows a steady decline from the early 1980s through the late 2010s, consistent with sustained improvements in longevity and health outcomes. The rate of decline accelerates during the 2000s and moderates in more recent years. The close alignment between the two series at different frequencies is expected, as both the annual mortality rates and the scaled monthly death rates are driven by the same underlying mortality trend. The monthly series simply retains additional intra-year variation that is averaged out in the annual rates. Additionally, we verify that the annual death counts used by HMD to compute annual mortality rates are approximately equal to the sum of monthly deaths from the CDC, with differences typically no greater than five deaths.

\section{State-space framework with mixed-frequency mortality data}
\label{sec:mfss_model}
\subsection{Lee-Carter as a state-space model}

The state-space model provides a flexible framework for representing dynamic systems such as time series processes. It consists of two components: a state equation governing the evolution of latent states over time, and an observation/measurement equation linking these latent states to the observed data.

Many mortality models can be expressed within this framework. The Lee--Carter model \citep{leeandcarter1992} for annual mortality rates,
\begin{align}
\ln m_{x,t} &= a_x + b_x k_t + \varepsilon_{x,t}, \label{eq:obs_eq}\\
k_t &= k_{t-1} + d + \eta_t,\label{eq:state_eq}
\end{align}
can be viewed naturally as a state-space model. Equation \eqref{eq:obs_eq} links the observed log mortality rate $\ln m_{x,t}$ to the latent period factor $k_t$, and equation \eqref{eq:state_eq} specifies $k_t$ as a random walk with drift.

Casting Lee--Carter-type models in state-space form has been explored in the literature and it enables likelihood-based estimation and forecasting using tools such as the Kalman filter/smoother and simulation-based methods; see, for example, \cite{pedroza2006,fung2015, Fung2017, Fung2019}.

In our application, mortality is observed at two temporal resolutions:  annual mortality rates and monthly deaths. To exploit both sources of information within a single framework, we extend the Lee–Carter specification to a mixed-frequency state-space model (MF-SS) that jointly models the annual and monthly series through a common monthly latent process.

\subsection{Notation}

Let $x=x_0,\ldots,x_1$ denote age, and let $t=t_0,\ldots,t_1$ denote calendar year. Within each year $t$, $(t,h)$ denotes month $h\in\{1,\ldots,12\}$.
Let $n_a=x_1-x_0+1$ be the number of ages. Define:
\begin{itemize}
    \item $P_{x,t}$: population at the beginning of year $t$ for age $x$.
    \item $d_{x,(t,h)}$: deaths at age $x$ in month $h$ of year $t$.
    \item $m_{x,t}$: annual mortality rate at age $x$ in year $t$, observed at year end.
    \item $y_{x,t}=\ln m_{x,t}$, and $\bm{y}_t=[y_{x_0,t},\ldots,y_{x_1,t}]^\top$.
    \item The log scaled monthly death  rate
    \[
    z_{x,(t,h)}=\ln\!\left(\frac{d_{x,(t,h)}}{P_{x,t}}\right),
    \qquad
    \bm{z}_{(t,h)}=[z_{x_0,(t,h)},\ldots,z_{x_1,(t,h)}]^\top.
    \]
    We scale monthly deaths by the start-of-year population so that variation in $z_{x,(t,h)}$ primarily reflects mortality variation rather than population changes. Note that $z_{x,(t,h)}$ is not a true monthly central death rate, for which the denominator would be a monthly exposure measure.\footnote{We do not model monthly central death rates directly because doing so would require estimating monthly exposures during model fitting, introducing an additional layer of measurement error and variability. Monthly exposures are nevertheless required at the forecasting stage when we convert predicted deaths into annual mortality rates. We approximate these exposures as described in Section \ref{sec:forecast_type}.}
\end{itemize}

\subsection{Constructing a state-space model for mixed-frequency mortality data}

We construct a mixed-frequency state-space (MF-SS) model that links annual and monthly mortality through a common monthly latent process. The model consists of two observation equations:
\begin{equation}
\label{eq:MF-SS_obs}
\begin{aligned}
y_{x,t} &= a_{1,x} + b_{1,x} k_{1,t} + \varepsilon_{1,x,t},\\
z_{x,(t,h)} &= a_{2,x} + b_{2,x} k_{2,(t,h)} + \varepsilon_{2,x,(t,h)},
\end{aligned}
\end{equation}
where $k_{1,t}$ is the annual period factor, $k_{2,(t,h)}$ is the latent monthly period factor at month $h$ of year $t$, $\varepsilon_{1,x,t}\sim N(0,\sigma_{1,x}^2)$ and $\varepsilon_{2,x,(t,h)}\sim N(0,\sigma_{2,x}^2)$ are independent across time and across ages. The parameters $a_{1,x}$ and $a_{2,x}$ capture age-specific mortality level in the annual and monthly series, while $b_{1,x}$ and $b_{2,x}$ measure age-specific sensitivities to the underlying mortality index. 

To connect annual and monthly components, we define the annual period factor as the intra-year average of the monthly factors,
\[
k_{1,t}=\frac{1}{12}\sum_{h=1}^{12}k_{2,(t,h)}.
\]
Because the annual series is a rate and the monthly series is a scaled count-based measure, allowing $b_{1,x}$ and $b_{2,x}$ to differ accommodates systematic differences between annual and monthly measurements, while still linking both series to the same latent driver.

For estimation and forecasting, it is convenient to index months by a single integer $\tau=1,\ldots,T$, with each $\tau$ corresponding to a unique $(t,h)$. We write the monthly latent factor as $k_{2,\tau}$.

We model $\{k_{2,\tau}\}$ using a seasonal ARIMA (SARIMA) process with drift. We considered a range of candidate SARIMA$(p,d,q)(P,D,Q)_{12}$ specifications for $k_{2,\tau}$ and selected orders using BIC. The selected specification for our application is a SARIMA$(2,0,0)(0,1,0)_{12}$ with drift, written as
\begin{equation}
\label{eq:MF-SS_state}
(1-\phi_1L-\phi_2L^2)(1-L^{12})\,k_{2,\tau}=\mu+w_\tau,
\qquad
w_\tau\sim N(0,\sigma_w^2),
\end{equation}
where $L$ is the lag operator.

As with most Lee-Carter-type models, the parameters in the observation equations are not uniquely identified. To ensure identifiability, we impose the following constraints:
\[
\sum_{\tau=1}^{T} k_{2,\tau}=0,
\qquad
\sum_{x = x_0}^{x_1} b_{1,x} = 1, \qquad 
\sum_{x=x_0}^{x_1} b_{2,x}=1.
\]

\subsection{The state-space form}
\label{sec:ss_form}

While the mixed-frequency Lee–Carter model can be specified in terms of the age- and time-specific equations in \eqref{eq:MF-SS_obs}–\eqref{eq:MF-SS_state}, it is convenient to rewrite the system in compact state-space form for estimation and forecasting. 
The state-space representation makes explicit that the model is linear and Gaussian and provides a unified framework for likelihood-based inference, for handling missing observations, and for extending the model (for example by adding further latent factors or additional data series) without changing the underlying structure.

The SARIMA$(2,0,0)(0,1,0)_{12}$ dynamics imply that $k_{2,\tau}$ depends on lags 1, 2, 12, 13 and 14:
\[
k_{2,\tau}
=
\mu + w_\tau
+ \phi_1 k_{2,\tau-1} + \phi_2 k_{2,\tau-2}
+ k_{2,\tau-12} - \phi_1 k_{2,\tau-13} - \phi_2 k_{2,\tau-14}.
\]
To express this as a first-order Markov system, define the 15-dimensional state vector
\[
\bm{k}_\tau = [k_{2,\tau},k_{2,\tau-1},\ldots,k_{2,\tau-14}]^\top.
\]
Then the latent factor dynamics can be expressed as
\begin{equation}
\label{eq:MF-SS_state_vector}
\bm{k}_\tau=\bm{H}\bm{k}_{\tau-1}+\bm{u}+\bm{G}w_\tau,
\end{equation}
where $\bm{u}=[\mu,0,\ldots,0]^\top$ and $\bm{G}=[1,0,\ldots,0]^\top$. The transition matrix $\bm{H}$ has companion form: its first row contains the coefficients on $k_{2,\tau-1},k_{2,\tau-2},k_{2,\tau-12},k_{2,\tau-13},k_{2,\tau-14}$ (with zeros elsewhere), and rows 2--15 shift the lags down by one. Specifically,
\[
    \bm{H} =
    \begin{bmatrix}
        \phi_1 & \phi_2 & 0 & \cdots & 0 & 1 & -\phi_1 & -\phi_2 & 0\\
        1 & 0 & \cdots & 0 & 0 & 0 & 0 & 0 & 0\\
        \vdots &  & \ddots &  &  &  &  &  & \vdots\\
        0 &  &  &  &  &  &  & 1 & 0
    \end{bmatrix}.
    \]

At month $\tau$, we stack annual and monthly observations as
\[
\bm{Y}_\tau=
\begin{bmatrix}
\bm{y}_\tau\\
\bm{z}_\tau
\end{bmatrix},
\]
where $\bm{z}_\tau=\bm{z}_{(t,h)}$ is observed when monthly data are available, and $\bm{y}_\tau$ equals $\bm{y}_t$ only when $\tau$ corresponds to December of year $t$ and is treated as missing otherwise. 

Similarly, we stack the observation errors. Let $\bm{\varepsilon}_{1,t}=[\varepsilon_{1,x_0,t},\ldots,\varepsilon_{1,x_1,t}]^\top$ and $\bm{\varepsilon}_{2,\tau}=[\varepsilon_{2,x_0,\tau},\ldots,\varepsilon_{2,x_1,\tau}]^\top$. Define $\bm{\varepsilon}_{1,\tau}=\bm{\varepsilon}_{1,t}$ only when $\tau$ corresponds to December of year $t$ and treat it as missing otherwise.
The stacked observation errors are
\[
\bm{\varepsilon}_\tau=
\begin{bmatrix}
\bm{\varepsilon}_{1,\tau}\\
\bm{\varepsilon}_{2,\tau}
\end{bmatrix},
\qquad
\bm{\varepsilon}_\tau\sim\mathcal{N}(\bm{0},\bm{R}),
\qquad
\bm{R}=
\begin{bmatrix}
\bm{R}_1 & 0\\
0 & \bm{R}_2
\end{bmatrix},
\]
with   $\bm{R}_1=\mathrm{diag}(\sigma^2_{1,x_0},\ldots,\sigma^2_{1,x_1})$ and $\bm{R}_2=\mathrm{diag}(\sigma^2_{2,x_0},\ldots,\sigma^2_{2,x_1})$.

Let $\bm{a}_1=[a_{1,x_0},\ldots,a_{1,x_1}]^\top$ and $\bm{a}_2=[a_{2,x_0},\ldots,a_{2,x_1}]^\top$. Define $\bm{A}=[\bm{a}_1^\top,\bm{a}_2^\top]^\top$, and $\bm{B}$ as a $2n_a \times 15$ loading matrix of the form
\[
\bm{B} =
\begin{bmatrix}
\frac{b_{1,x_0}}{12} & \cdots & \frac{b_{1,x_0}}{12} & 0 & \cdots & 0 \\
\vdots                    &        & \vdots                    & \vdots &        & \vdots \\
\frac{b_{1,x_1}}{12} & \cdots & \frac{b_{1,x_1}}{12} & 0 & \cdots & 0 \\
b_{2,x_0}            & 0      & \cdots & 0                & 0 & \cdots & 0 \\
\vdots                    & \vdots &        & \vdots           & \vdots & \ddots & \vdots \\
b_{2,x_1}            & 0      & \cdots & 0                & 0 & \cdots & 0
\end{bmatrix},
\]
so that the annual equation loads equally on the 12 months of the calendar year through $\frac{b_{1,x}}{12}$, while the monthly equation loads on the current monthly factor through $b_{2,x}$.

With this notation, the mixed-frequency model can be written as
\begin{align}
\bm{Y}_\tau &= \bm{A} + \bm{B}\bm{k}_\tau + \bm{\varepsilon}_\tau, \label{eq:MF-SS_stacked_obs}\\
\bm{k}_\tau &= \bm{H}\bm{k}_{\tau-1} + \bm{u} + \bm{G}w_\tau. \label{eq:MF-SS_stacked_trans}
\end{align}
This compact state-space representation accommodates both annual and monthly mortality data on a common monthly time grid and allows us to develop the likelihood-based estimation and forecasting procedures introduced in the following sections. Equations \eqref{eq:MF-SS_stacked_obs} and \eqref{eq:MF-SS_stacked_trans} are also referred to as observation/measurement equation and state/transition equation respectively.

\subsection{Mixed-frequency state-space model estimation}
\label{sec:fitting}

Let $\bm{\Theta}$ denote the collection of parameters in the MF-SS model,
\[
\bm{\Theta}=\{\bm{A},\bm{B},\bm{H},\bm{R},\bm{u},\sigma_w^2\}.
\]
We estimate $\bm{\Theta}$ by maximum likelihood using an Expectation--Maximisation (EM) algorithm. The EM approach is convenient because the latent monthly state sequence $\{\bm{k}_\tau\}_{\tau=1}^{T}$ enters the likelihood through both the observation and state equations. In the E-step, we compute conditional moments of $\bm{k}_{1:T}$ given the observed data via the Kalman filter and smoother; in the M-step, we update parameters by maximising the expected complete-data log-likelihood.

First introduced by \citet{kalmanRE1960}, the Kalman filter and smoother is a fundamental tool for estimating latent states in state-space models. It has become standard in economic and demographic time-series analysis, providing a systematic way to infer unobserved processes that evolve over time. The method extends naturally to mixed-frequency settings, where low-frequency observations are treated as missing at higher-frequency time points. This ``missing observations'' treatment is widely used in the mixed-frequency literature (e.g., \citealp{Mariano2002}).

In our mixed-frequency system, $\bm{Y}_\tau$ stacks monthly and annual components, but the annual component is observed only at calendar year-end. Therefore, for intra-year months, the annual entries are treated as missing. The complete observation vector  at each month $\tau$ can be partitioned as
\[
\bm{Y}_\tau=
\begin{bmatrix}
\bm{Y}^{\mathrm{obs}}_\tau\\
\bm{Y}^{\mathrm{mis}}_\tau
\end{bmatrix},
\]
where $\bm{Y}^{\mathrm{obs}}_\tau$ represents the observed components and $\bm{Y}^{\mathrm{mis}}_\tau$ denotes the missing annual entries in intra-year months.
In intra-year months, the missingness is handled directly within the Kalman filter and smoother by conditioning only on  monthly information $\bm{Y}^{\mathrm{obs}}_\tau$. At year-end, both annual and monthly components are included in the filtering and smoothing. More details on the ``missing observations'' treatment in state-space EM can be found in \citet{elizabeth2013}. 

At iteration $i$, the EM algorithm alternates between:
\begin{itemize}
    \item \textbf{E-step (state estimation).}
    Given current parameter estimates $\bm{\Theta}^{(i)}$, we compute the filtering and smoothing distributions of the latent states $\{\bm{k}_\tau\}$ given the observed mortality series. 
    The Kalman filter proceeds forward in time through a two-step recursion:
    \begin{itemize}
        \item Prediction step. Using the state equation and the filtered state from month $\tau-1$, the model forms the one-step-ahead (prior) distribution of the state at month $\tau$,
  \[
  (\bm{k}_\tau \mid \bm{Y}^{\mathrm{obs}}_{1:\tau-1},\bm{\Theta}^{(i)})
  \;\sim\;
  \mathcal{N}\!\big(\bm{k}_{\tau|\tau-1},\,\bm{P}_{\tau|\tau-1}\big),
  \]
  where $\bm{k}_{\tau|\tau-1}$ and $\bm{P}_{\tau|\tau-1}$ denote the predicted mean and covariance, respectively, and their detailed expressions are provided in Appendix \ref{app:kf_update}. This prior implies a forecast distribution for the new observation $\bm{Y}^{\mathrm{obs}}_\tau$ under the observation equation, and hence a forecast mean and forecast covariance for $\bm{Y}^{\mathrm{obs}}_\tau$.

        \item Update step. Upon observing $\bm{Y}^{\mathrm{obs}}_\tau$, the filter computes the one-step-ahead prediction error (innovation),
  \[
  \bm{v}_\tau
  =
  \bm{Y}^{\mathrm{obs}}_\tau
  - \mathbb{E}\!\left[\bm{Y}^{\mathrm{obs}}_\tau \mid \bm{Y}^{\mathrm{obs}}_{1:\tau-1},\bm{\Theta}^{(i)}\right],
  \]
  together with its covariance $\bm{S}_\tau=\mathrm{Var}(\bm{v}_\tau\mid \bm{Y}^{\mathrm{obs}}_{1:\tau-1},\bm{\Theta}^{(i)})$.
  The Kalman gain $\bm{K}_\tau$ is then formed from $\bm{P}_{\tau|\tau-1}$ and $\bm{S}_\tau$, and the prediction error is used to update the distribution of the state: 
  \[  (\bm{k}_\tau \mid \bm{Y}^{\mathrm{obs}}_{1:\tau},\bm{\Theta}^{(i)})
\sim
\mathcal{N}\!\big(\bm{k}_{\tau|\tau},\,\bm{P}_{\tau|\tau}\big),\]
 where $
  \bm{k}_{\tau|\tau}
  =
  \bm{k}_{\tau|\tau-1} + \bm{K}_\tau \bm{v}_\tau$ and $
  \bm{P}_{\tau|\tau}
  =
  \mathrm{Var}(\bm{k}_\tau\mid \bm{Y}^{\mathrm{obs}}_{1:\tau},\bm{\Theta}^{(i)}).
$

  Intuitively, $\bm{K}_\tau$ downweights the prediction error when observation noise is large and upweights it when the model forecast is more uncertain. When the annual components of $\bm{Y}_\tau$ are missing in intra-year months, the update conditions only on the observed component $\bm{Y}^{\mathrm{obs}}_\tau$ and ignores the missing entries. The detailed expressions for  $\bm{K}_\tau$, $\bm{P}_{\tau|\tau}$ and $\bm{S}_\tau$ are provided in Appendix \ref{app:kf_update}.
\end{itemize}

    After the forward filtering pass, the Kalman smoother runs backward through time to compute the smoothing distribution (posterior)
    $
    p(\bm{k}_{1:T}\mid \bm{Y}^{\mathrm{obs}}_{1:T}, \bm{\Theta}^{(i)}).$
    In the linear-Gaussian case, this distribution is Gaussian and is therefore fully characterised by its first- and second-order moments. The smoother returns the smoothed state means
    \[
    \widetilde{\bm{k}}_\tau
    =
    \mathbb{E}\!\left[\bm{k}_\tau \mid \bm{Y}^{\mathrm{obs}}_{1:T}, \bm{\Theta}^{(i)}\right],
    \]
    together with $\mathbb{E}[\bm{k}_\tau\bm{k}_\tau^\top\mid \bm{Y}^{\mathrm{obs}}_{1:T}]$ and
$\mathbb{E}[\bm{k}_\tau\bm{k}_{\tau-1}^\top\mid \bm{Y}^{\mathrm{obs}}_{1:T}]$, which constitute the sufficient statistics needed for the M-step.

\item \textbf{M-step (parameter updating).}
We update the parameters by maximising the expected complete-data log-likelihood,
\[
Q(\bm{\Theta}\mid \bm{\Theta}^{(i)})
=
\mathbb{E}\!\left[
\log f(\bm{Y}_{1:T},\bm{k}_{1:T}\mid \bm{\Theta})
\ \big|\ 
\bm{Y}^{\mathrm{obs}}_{1:T},\bm{\Theta}^{(i)}
\right].
\]
Here, $f(\bm{Y}_{1:T},\bm{k}_{1:T}\mid \bm{\Theta})$ denotes the joint density of the complete data $(\bm{Y}_{1:T},\bm{k}_{1:T})$ implied by the state-space model. Under the linear--Gaussian assumptions, this joint density is multivariate Gaussian. The expectation in $Q(\bm{\Theta}\mid \bm{\Theta}^{(i)})$ is taken with respect to the joint conditional distribution of the unobserved quantities, including the latent states and the annual mortality missing for intra-year months, 
\[
(\bm{k}_{1:T},\bm{Y}^{\mathrm{mis}}_{1:T}) \mid \bm{Y}^{\mathrm{obs}}_{1:T},\bm{\Theta}^{(i)}.
\]
In the linear--Gaussian case, $Q(\bm{\Theta}\mid \bm{\Theta}^{(i)})$ is quadratic in the parameters, so the maximisation yields closed-form updates based on a set of expected sufficient statistics.
\end{itemize}

These two steps are iterated until convergence. In practice, we monitor both the increase in the $Q$-function and the observed-data log-likelihood
\(
\ell(\bm{\Theta}) = \log f(\bm{Y}_{1:T}^{\mathrm{obs}} \mid \bm{\Theta}),
\)
and stop when the relative change in these quantities falls below a pre-specified threshold. Detailed expressions for the complete-data log-likelihood, the $Q$-function, and the observed-data log-likelihood are provided in Appendix \ref{app:complete_ll}, \ref{app:Q_func} and \ref{app:obs_ll}, respectively. The estimation algorithm is summarised in Appendix \ref{app:algorithm}. Further details on Kalman filtering and smoothing can be found in \cite{kalmanRE1960,koopman2012}.

\subsection{Forecasting the latent factor and future mortality}
A key advantage of the state-space representation is that, conditional on estimated parameters, the latent state can be updated as new observations become available without re-estimating the model. In our implementation, parameter re-estimation is performed only after a full additional calendar year of data has been observed. Within the year, we hold the parameter vector fixed at its most recent estimate. This strategy allows us to incorporate new information in a timely manner while avoiding frequent re-estimation and the resulting parameter volatility.

Suppose the model is estimated using data up to December of year $t$, yielding the parameter estimate $\widehat{\bm{\Theta}}$. Let $\tau$ index months sequentially and let $T$ denote the end of the fitting period, December of year $t$. We initialise forecasting from the end-of-sample state distribution. Denote by $(\bm{k}_{T|T},\bm{P}_{T|T})$ the filtered state mean and covariance at time $T$, which coincides with the smoothed state at $T$.

Consider year $t+1$, and suppose monthly mortality observations are available for months $1,\ldots,h$ (with $h\leq 12$). For each month $\tau=T+1,\ldots,T+h$, we keep the parameter estimate $\widehat{\bm{\Theta}}$ unchanged and update the latent state by iterating the standard Kalman prediction--update recursion under $\widehat{\bm{\Theta}}$:
\begin{itemize}
  \item Prediction step: propagate $(\bm{k}_{\tau-1|\tau-1},\bm{P}_{\tau-1|\tau-1})$ through the state equation to obtain the prediction distribution $\mathcal{N}(\bm{k}_{\tau|\tau-1},\bm{P}_{\tau|\tau-1})$.
  \item Update step: condition on the newly observed monthly data to obtain the filtered distribution $\mathcal{N}(\bm{k}_{\tau|\tau},\bm{P}_{\tau|\tau})$.
\end{itemize}
Because the annual observation is not available within the year, $\bm{Y}_\tau$  contains missing components in months $1,\ldots,11$. The update step therefore conditions only on the observed monthly entries, consistent with the missing-observation treatment described in Section \ref{sec:fitting}. 

For horizons beyond the observed window, i.e., for $\tau=T+1,T+h+2,\ldots$, no additional observations are available and the latent state is propagated forward using the prediction step only. This yields the multi-month-ahead predictive distribution of the state,
$(\bm{k}_{\tau|T+h},\bm{P}_{\tau|T+h})$,
under the state equation and $\widehat{\bm{\Theta}}$. Forecasts of monthly and annual mortality are then obtained by substituting the state forecasts into the observation equations.

As additional months of data in year $t+1$ become available, the above procedure extends seamlessly. Each new observation adds one Kalman update step, thereby shortening the purely predictive horizon and refining the latent state estimate in real time, without re-running the EM algorithm.

Although the predictive distribution of the latent state is available in closed form under linear-Gaussian assumptions, the forecast target (e.g. annual mortality rates) may involve nonlinear transformations. We therefore approximate the predictive distribution of the forecast targets via Monte Carlo simulation.
For months $1,\ldots,h$ of year $t+1$, the monthly observations are incorporated via the Kalman filter update, yielding the filtered state in closed form under the linear–Gaussian specification. Simulation is only used to approximate predictive distributions for unobserved future months and any nonlinear transformations of the forecast targets.

Specifically, at the forecast origin, $T+h$, where the observations for first $h$ months of year $t+1$ are available, we generate $B=10{,}000$ sample paths from the fitted state-space model. For each draw $b=1,\ldots,B$, we obtain a future path of the latent state $\{\bm{k}^{(b)}_{\tau}\}$ by simulating its error terms and using the state equation, and then simulate the observation errors in the observation equations. These simulated states and errors are mapped to annual mortality rates and monthly scaled death probabilities using the observation equations. Point forecasts are reported as Monte Carlo means, and $(1-\alpha)$ prediction intervals are obtained from the empirical $\alpha/2$ and $1-\alpha/2$ quantiles of the simulated predictive distribution.




\section{Independent annual and monthly Models}
\label{sec:ind_LC}
A natural alternative for generating both annual and monthly forecasts is to model the two series separately, for example, by fitting a Lee--Carter (LC) model to the annual mortality series and another LC model to the scaled monthly series. This approach is simple, but it does not guarantee that the annual forecasts implied by aggregating monthly predictions align with the annual LC forecasts. Forecast reconciliation can be applied to these independently generated forecasts so that information from both temporal resolutions is incorporated. We discuss the reconciliation implementation in Section~\ref{sec:reconciliation}.

For annual mortality, we consider a standard Lee-Carter model \citep{leeandcarter1992}: 
\begin{align*}
\ln m_{x,t} &= a_{3,x} + b_{3,x} k_{3,t} + \varepsilon_{3,x,t},
\qquad \varepsilon_{3,x,t} \sim \mathrm{MVN}\!\left(\bm{0}, \bm{\Sigma}_3\right),\\
k_{3,t+1} &= k_{3,t} + d + \eta_{3,t},
\qquad \eta_{3,t} \sim N(0, \sigma_{\eta_3}^2),
\end{align*}
where $a_{3,x}$ captures the age-specific annual mortality level, $b_{3,x}$ is the age-specific sensitivity on the period factor $k_{3,t}$, and $\bm{\Sigma_3}$ is diagonal covariance matrix. The period factor follows a random walk with drift, which captures the long-run trend in mortality improvements. We also considered higher-order ARIMA specifications for $k_{3,t}$, but selected the random walk with drift based on BIC. To ensure identifiability, we impose the standard constraints $\sum_x b_{3,x} = 1$ and $\sum_t k_{3,t} = 0$.

For the scaled monthly series, we fit a LC model at the monthly index $\tau$,
\begin{align*}
&z_{x,\tau} = a_{4,x} + b_{4,x} k_{4,\tau} + \varepsilon_{4,x,\tau},
\qquad \varepsilon_{4,x,\tau} \sim \mathrm{MVN}\!\left(\bm{0}, \bm{\Sigma_4}\right),\\
&(1 - \varphi_1 L - \varphi_2 L^2)(1 - L^{12})\, k_{4,\tau}
= c + \eta_{4,\tau},
\qquad 
\eta_{4,\tau} \sim N(0,\sigma_{\eta_4}^2),
\end{align*}
where $a_{4,x}$ and $b_{4,x}$ play the same roles as in the annual model, and $\bm{\Sigma}_4$ is again assumed diagonal. We specify the monthly period factor $k_{4,\tau}$ to follow the same SARIMA structure used for the latent factor in the state-space model, ensuring comparability in the assumed time-series dynamics. Identifiability is enforced via $\sum_x b_{4,x} = 1$ and $\sum_{\tau} k_{4,\tau} = 0$.
Parameters in both models are estimated by maximum likelihood under Gaussian errors, consistent with the distributional assumptions used in the state-space framework. 


As intra-year monthly observations accumulate, the annual LC model remains unchanged because it is estimated from complete annual values. Consequently, neither $(a_{3,x},b_{3,x})$ nor the annual period factor $k_{3,t}$ can be updated until the end-of-year observation is finalised. By contrast, the monthly LC model can be updated as each new month becomes available. Consistent with our treatment of intra-year information arrival, we keep the previously estimated age profiles $(\hat a_{4,x},\hat b_{4,x})$ fixed and update only the current period factor when a new month is observed.
Other model parameters are re-estimated only once a full additional calendar year of data becomes available.

Specifically, when a new month $\tau$ is observed, we estimate $\hat{k}_{4,\tau}$ by a no-intercept least-squares regression across ages,
\[
z_{x,\tau}-\hat a_{4,x} = \hat b_{4,x}\, k_{4,\tau} + \text{error}_x,
\]
so that
\begin{equation}
\hat{k}_{4,\tau}
=
\frac{\sum_x \hat b_{4,x}\bigl(z_{x,\tau}-\hat a_{4,x}\bigr)}{\sum_x \hat b_{4,x}^2}.
\label{eq:k_update_ols}
\end{equation}
The updated sequence $\{\hat k_{4,1:\tau}\}$ is then used as the most recent history for the SARIMA recursion, which produces revised forecasts for the remaining months of year without re-estimating model parameters.

Simulation is used only to obtain predictive distributions for future unobserved months and derived annual quantities. Specifically, we generate Monte Carlo paths of the future period factors from the fitted time-series models and draw observation errors from the Gaussian distributions. The simulated paths are substituted into the LC equations to obtain predictive distributions for the forecast targets, e.g., annual mortality rates. Since Monte Carlo forecasting for ARIMA and Lee--Carter models is standard, we omit further computational details.

\section{Estimation results and forecast evaluation}
\label{sec:results}
\subsection{Parameter estimates}\label{sec:in_sample_fit}

\begin{figure}[H]
    \centering
    \caption{Fitted age factor \(\hat{a}_{i,x}\) and age-specific loadings \(\hat{b}_{i,x}\) for SS and independent LC models, using data from 1999 to 2019 for ages 20-90.}
    \includegraphics[width=0.49\textwidth]{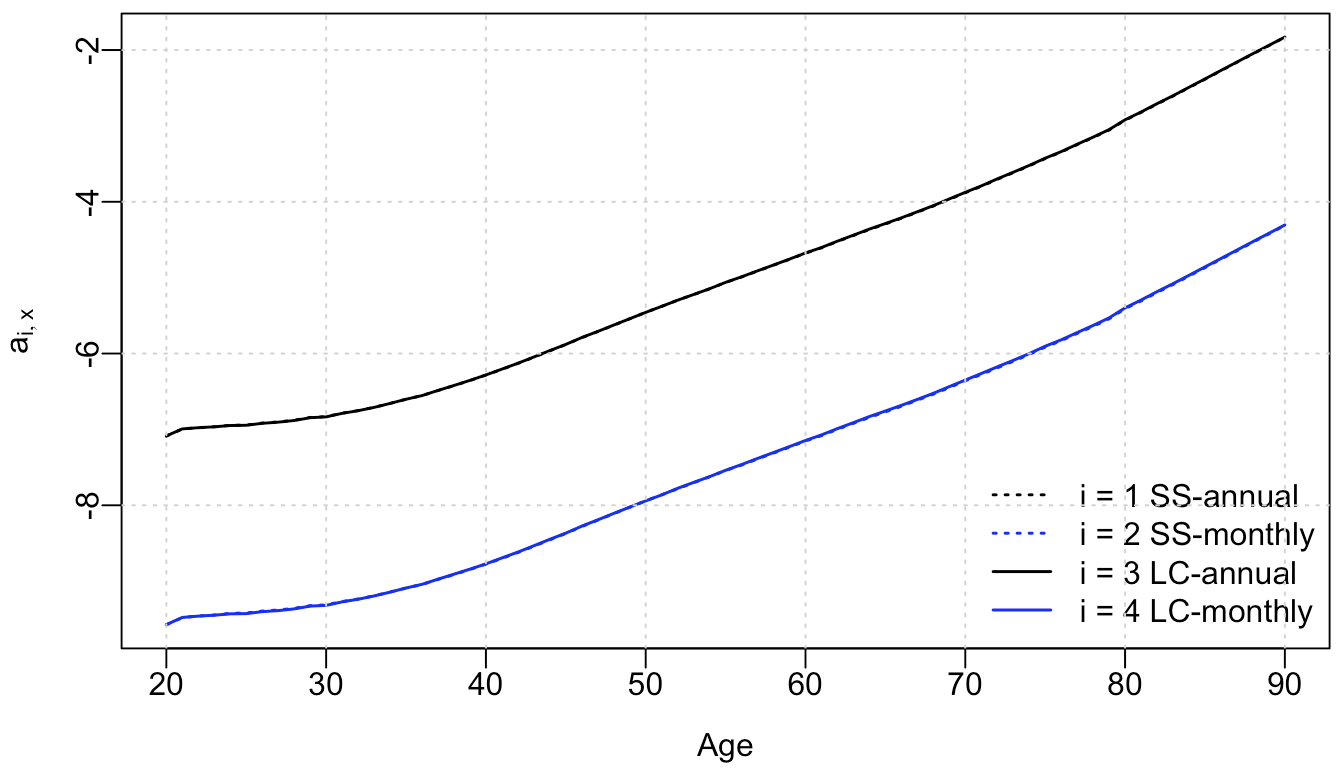}
    \includegraphics[width=0.49\textwidth]{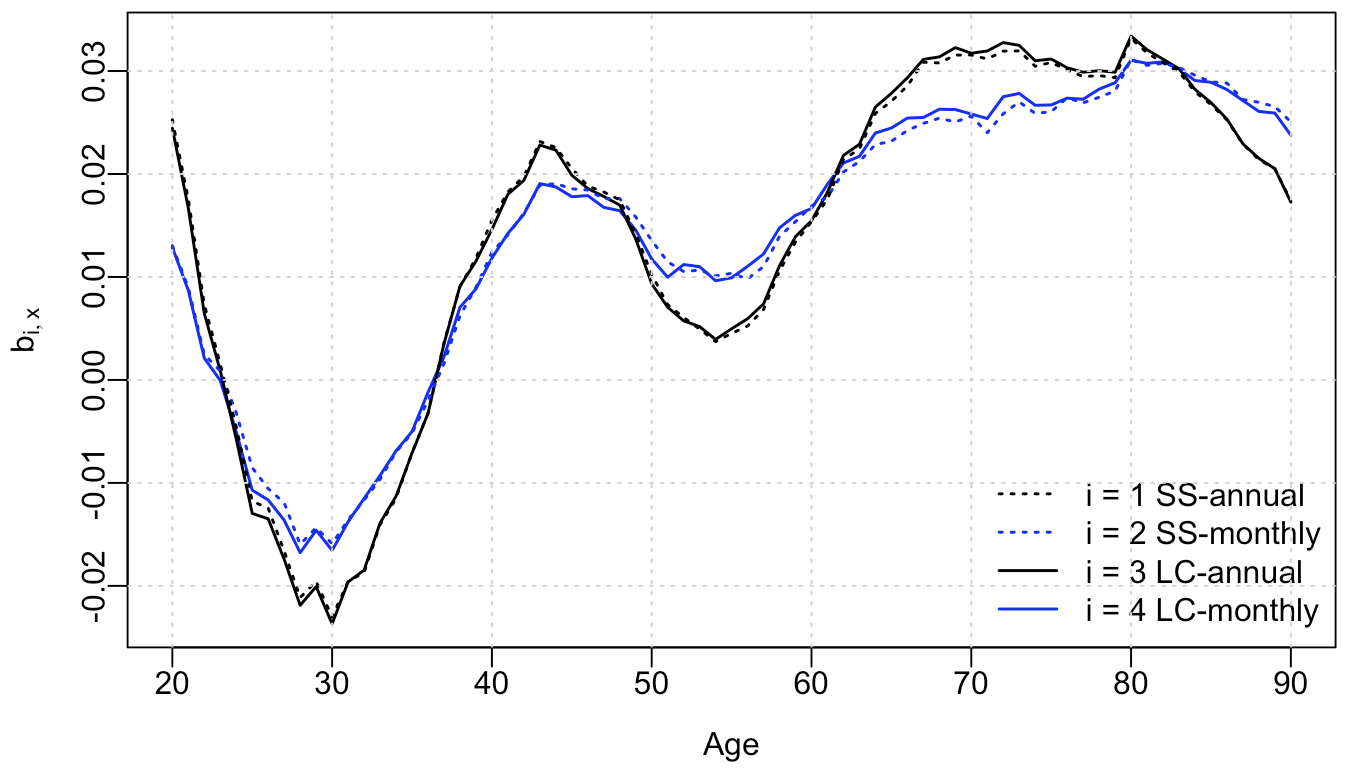}
    
    \label{fig:fitted_axbx}
\end{figure}

Figures~\ref{fig:fitted_axbx} compares the estimated age factor $a_{i,x}$ and age-specific sensitivity $b_{i,x}$ from the MF-SS model and the independent LC models, for both the annual and monthly specifications. The estimation uses the annual and monthly data from January 1999 to December 2019 for individuals aged 20-90. Hereafter, we use state-space (SS) model to refer to the mixed-frequency state-space (MF-SS) model. Overall, the SS and LC estimates exhibit very similar shapes, indicating that the SS model preserves the familiar Lee-Carter structure while integrating information across frequencies.

Within each frequency, the SS and LC estimates of the age factor $\hat{a}_{i,x}$ are nearly indistinguishable, suggesting that the baseline age-specific mortality level is robust to the modelling framework. The monthly estimates lie below the annual estimates across ages, reflecting the different scale of the monthly series relative to the annual mortality series.

The estimated $\hat{b}_{i,x}$ quantifies how sensitively mortality at each age responds to movements in the period factor. Again, SS and LC estimates are close within each frequency. Differences are more obvious when comparing monthly and annual sensitivities, indicating that the two monthly and annual models attribute somewhat different age patterns to their corresponding period factors.

\begin{figure}[H]
    \centering
    \caption{Fitted period factor \(\hat{k}_{i,\tau}\)  for SS and independent LC models, using mortality from 1999 to 2019 for ages 20-90.}
    \includegraphics[width=0.8\textwidth]{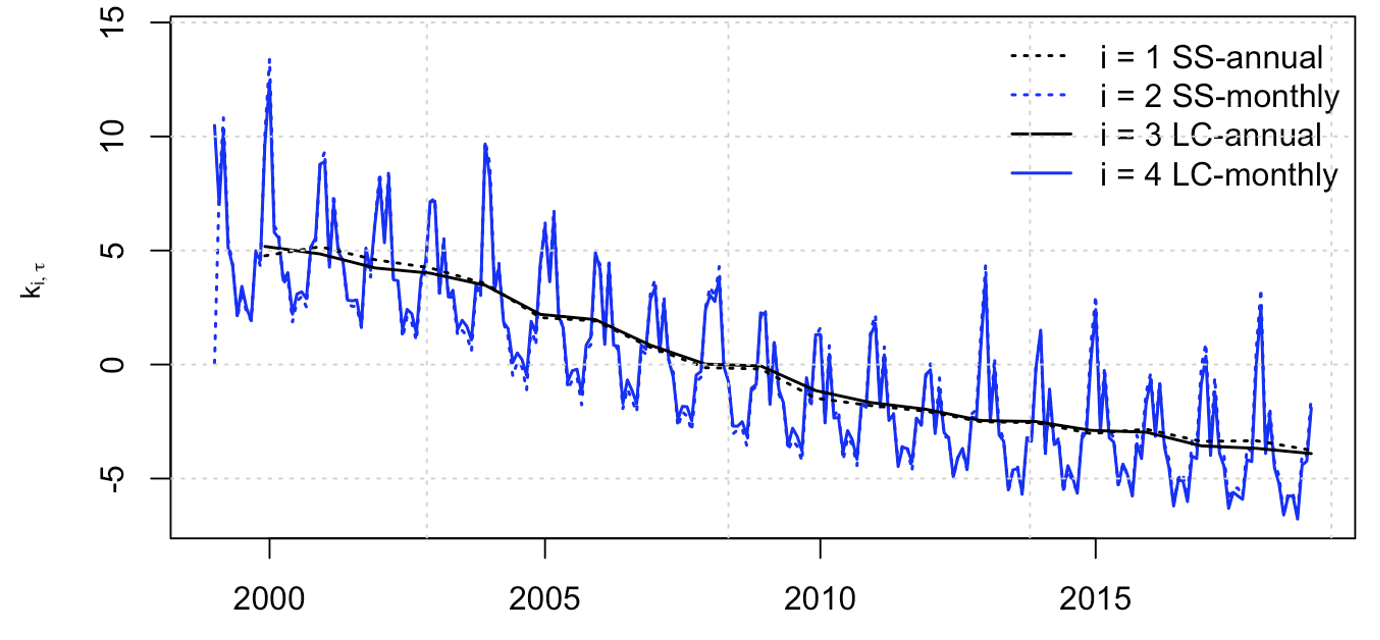}
    
    \label{fig:fitted_kt}
\end{figure}

Figure~\ref{fig:fitted_kt} shows the estimated period factors $\hat{k}_{i,\tau}$. The monthly factors exhibit pronounced short-run oscillations consistent with seasonality, whereas the annual factors are smooth and capture the long-run decline in mortality over the sample. The SS monthly factor fluctuates slightly more than the LC monthly factor, suggesting that the SS specification attributes a slightly stronger seasonal component to the data.

Differences between SS and LC are most visible at the start of the sample. This mainly reflects a state-space ``warm-up'' effect. Kalman filtering/smoothing estimates latent states sequentially, and early on there is limited history to pin down the seasonal component, so the initial estimates can be less stable. After roughly one seasonal cycle, the filter/smoother has accumulated enough information to produce more stable estimates of $k_{2,\tau}$. We therefore exclude the first 14 periods of $k_{2,\tau}$ from the E-step after applying identification constraints and base the M-step on the remaining, more stable portion of the sample.

\begin{figure}[H]
    \centering
    \caption{Average Kalman gain by age for the annual and monthly specifications, averaged across the fitting period 1999--2019.}
    \includegraphics[width=0.8\textwidth]{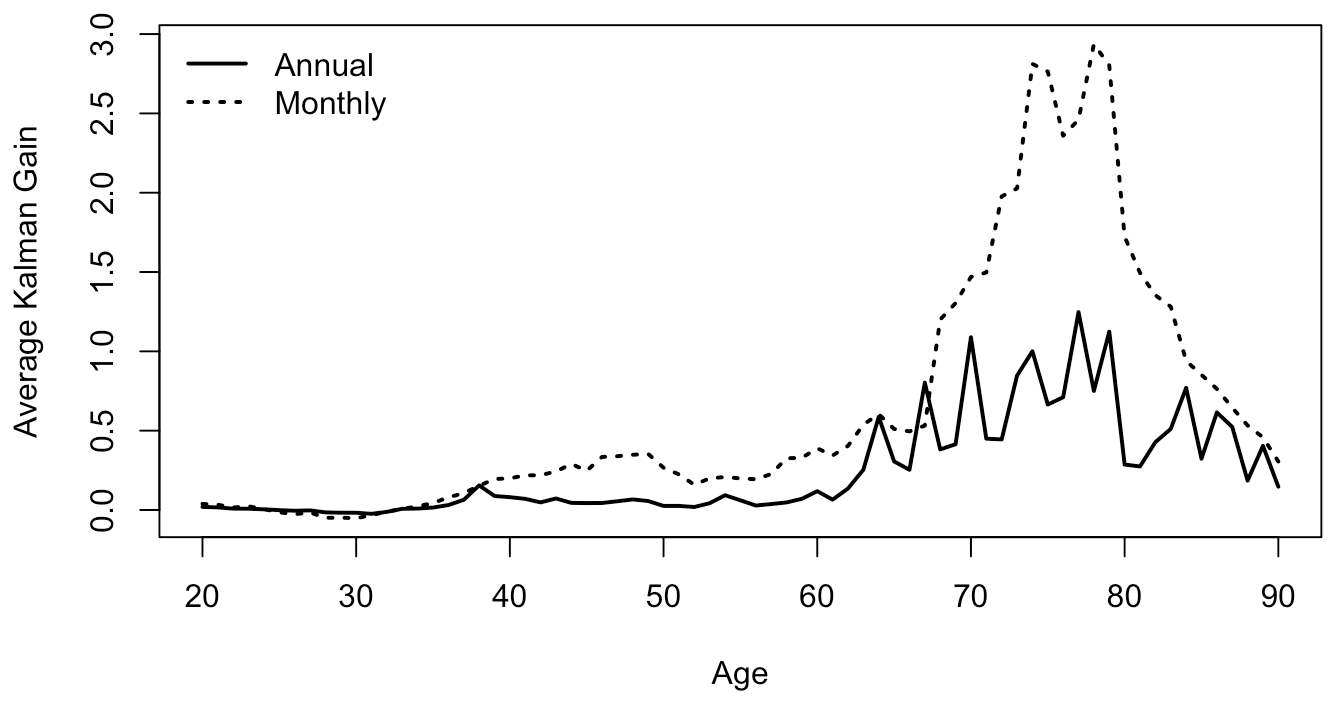}
    \label{fig:kalman_gain_comparison}
\end{figure}

Recall that the Kalman gain in the E-step quantifies how strongly the filter updates the latent state using the new observation, relative to relying on the one-step-ahead prediction from the model. A higher Kalman gain indicates that the observation  carries relatively more information compared with the model forecast, and therefore receives greater weight in the update.
Figure~\ref{fig:kalman_gain_comparison} shows the Kalman gain for monthly and annual observations by age averaged over 1999-2019. 
The monthly gains (dashed line) are consistently higher than the annual gains (solid line), meaning that the filter tends to adjust the latent factor more when a new monthly observation arrives than when a new annual observation arrives. This difference arises because the annual series aggregates monthly outcomes and is therefore smoother. Relative to the monthly signal, it typically contains less short-run variation. Consequently, the filter places less weight on the annual observation and more weight on the monthly observation when updating the monthly latent factor.

Both gain profiles increase with age, with the monthly gains rising most sharply around ages 70--80. This means that when new data arrive, the filter learns more from mortality at older ages and uses it more heavily to update the period factor. By contrast, observations at younger ages have little influence on the update, so they receive much smaller weights. 

\subsection{Mortality forecast evaluation}
\label{sec:forecasting}
\subsubsection{Direct and aggregated forecasts}
\label{sec:forecast_type}
 
Both the mixed-frequency state-space model and the independent Lee--Carter models can produce forecasts at annual and monthly horizons. In our setting, this naturally leads to three forecast targets:
\begin{enumerate}
    \item {Direct annual forecasts:} generated from the annual component of the model.
    \item {Direct monthly forecasts:} generated from the monthly component, capturing short-run movements and seasonal dynamics.
    \item {Aggregated monthly forecasts:} obtained by aggregating the monthly forecasts into an annual quantity, providing an annual projection implied by the higher-frequency dynamics. 
\end{enumerate}

In this paper, our primary interest is annual mortality rates. We therefore focus on two annual forecast objects: direct annual forecasts and aggregated monthly forecasts. Because these are produced through different observation equations, they need not coincide. In this section, we analyse them separately. Section~\ref{sec:reconciliation} then introduces temporal forecast reconciliation to combine information across the annual and monthly levels.

In both the state-space and independent LC models, the monthly observation is constructed as monthly deaths divided by the population at the beginning of the year. To forecast aggregated monthly rates, we use Monte Carlo simulation, the details of which are provided in Appendix \ref{sec:annualise}. Given a simulated monthly path $\{\tilde{z}_{x,(t,h)}\}_{h=1}^{12}$, we compute simulated monthly deaths
$\tilde{D}_{x,(t,h)} = P_{x,t}\exp(\tilde{z}_{x,(t,h)})$ and approximate monthly exposures $\tilde{E}_{x,(t,h)}$
using an updating procedure for population size within the year, as described in Appendix~\ref{sec:annualise}.
The implied annual mortality rate is then calculated as
\[
\tilde{m}_{x,t}=\frac{\sum_{h=1}^{12}\tilde{D}_{x,(t,h)}}{\sum_{h=1}^{12}\tilde{E}_{x,(t,h)}}.
\]

\subsection{Evaluating forecast accuracy}
\label{sec:eval_forecast}
\subsubsection{Forecast accuracy measure}
To illustrate how intra-year information and the use of both annual and monthly sources affect forecast accuracy, we evaluate forecast accuracy of four approaches: direct annual forecasts from the state-space model and the annual Lee--Carter model ($i=1$ SS--annual, $i=3$ LC--annual), and aggregated forecasts obtained from monthly models ($i=2$ SS--monthly, $i=4$ LC--monthly).

Let $\hat m^{(i)}_{x,t-1+n|(t,h)}$ denote the $n$-year-ahead forecast produced by method $i$ using information available up to month $h$ in year $t$. The absolute percentage error (APE) for the forecast is defined as
\[
\mathrm{APE}^{(i)}_{x,n|(t,h)} = \left|\frac{m_{x,t-1+n}-\hat m^{(i)}_{x,t-1+n|(t,h)}}{m_{x,t-1+n}}\right|,
\]
and the corresponding mean absolute percentage error (MAPE) across ages is 
\[
\mathrm{MAPE}^{(i)}_{n|(t,h)}=\frac{1}{n_a}\sum_{x\in [x_0,x_1]}\mathrm{APE}^{(i)}_{x,n|(t,h)}.
\]
We report results for $h\le 11$, since by $h=12$ the annual total is fully observed and the nowcast becomes trivial.

We perform both single-window and expanding-window analyses. 
In the single-window analysis, we fix a single forecast target year, e.g., $\mathcal{T}=2015$. For an $n$-year-ahead forecast, models are estimated once using all data available up to the end of year $\mathcal{T}-n$. In the forecast-origin year $t=\mathcal{T}-n+1$, we then generate intra-year forecasts at months $h=1,\ldots,11$ using information available up to month $h$, and evaluate forecast accuracy using $\mathrm{MAPE}^{(i)}_{n|(t,h)}$.

We also consider an expanding-window design with five forecast target years $\mathcal{T}\in\{2015,\ldots,2019\}$. We compute the average MAPE across target years as
\begin{equation}\label{eq:MAPE}
\mathrm{MAPE}^{(i)}_{n,h}=\frac{1}{|\mathcal{T}|}\sum_{\mathcal{T}=2015}^{2019}\mathrm{MAPE}^{(i)}_{n|(\mathcal{T}-n+1,h)},
\end{equation}
where $|\mathcal{T}|$ denotes the dimension of ${\mathcal{T}}$.

\subsubsection{Nowcast performance}
First, we focus on the one-year-ahead case ($n=1$). This is equivalent to a current year nowcast. We begin with a single forecast target year of 2015. For this 2015 illustration, all models are estimated using data from January 1999 to December 2014.
At the end of month $h\in\{1,\dots,11\}$ of 2015, we form an updated nowcast for the annual mortality rates in 2015. For the monthly approaches (SS--monthly and LC--monthly), the annual nowcast is constructed by combining the realised deaths counts for months $1,\dots,h$ with model-based projections for months $h+1,\dots,12$. 

The LC-annual forecast relies only on information available up to the end of 2014 and therefore provides the same annual forecast for all $h$.
In contrast, SS--annual is obtained from the joint state-space model with both annual and monthly observations. As monthly observations arrive within 2015, the Kalman filter updates the latent monthly factor, and the implied annual factor is updated through the structural constraint that the annual factor equals the average of the monthly factors. Hence SS--annual does incorporate intra-year monthly information. However, it incorporates this information through the latent-state update, rather than directly anchoring the annual forecast to the cumulative realised monthly totals as in the SS-monthly. This distinction is important for interpreting the forecast errors below.

\begin{figure}[H]
    \centering
    \caption{Left: Nowcast MAPE for forecast year 2015. Right: Average nowcast MAPE across expanding windows with forecast years 2015–2019. }
    \label{fig:forecast_one_step}\includegraphics[width=0.49\textwidth]{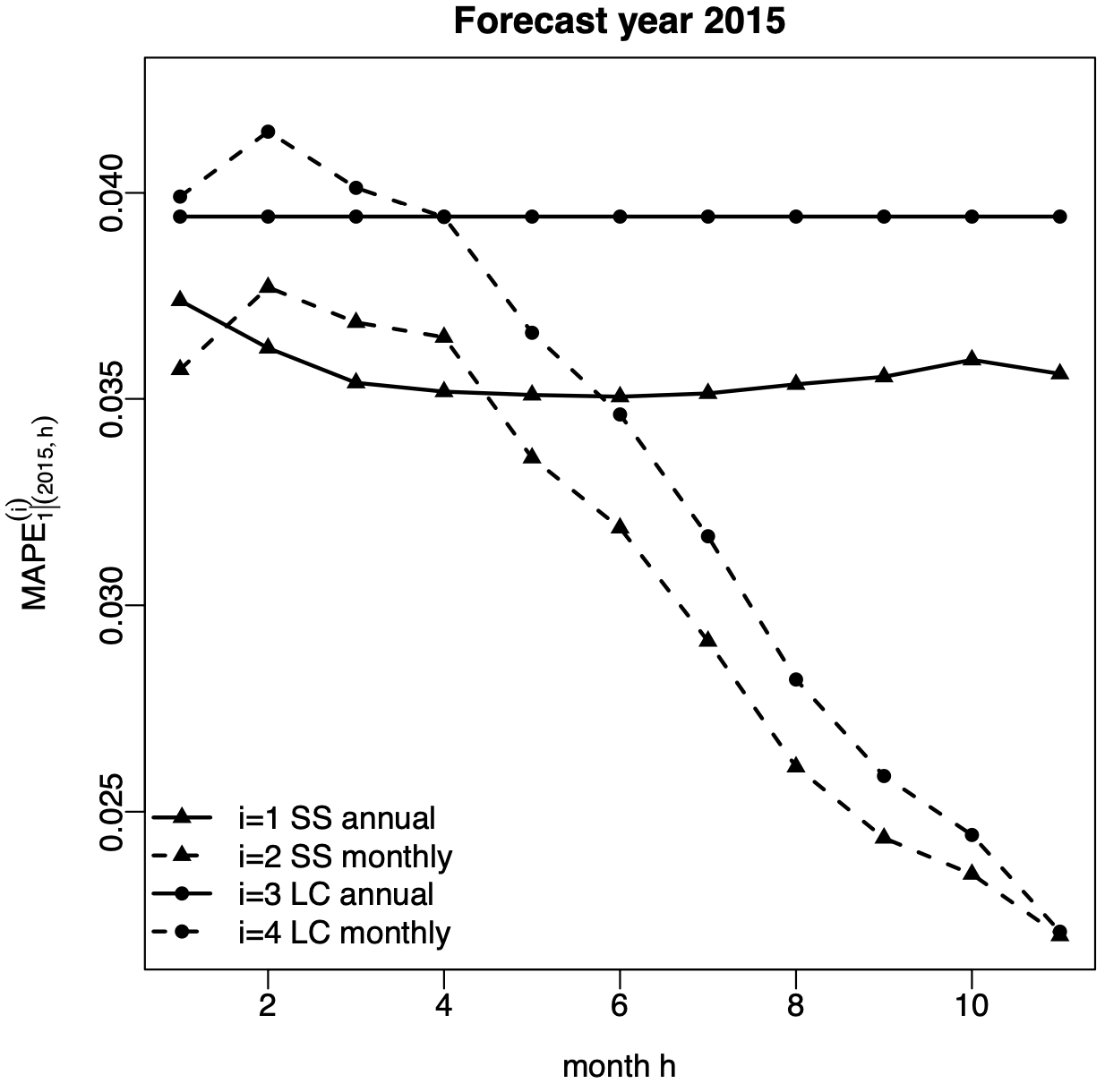}
    \includegraphics[width=0.49\textwidth]{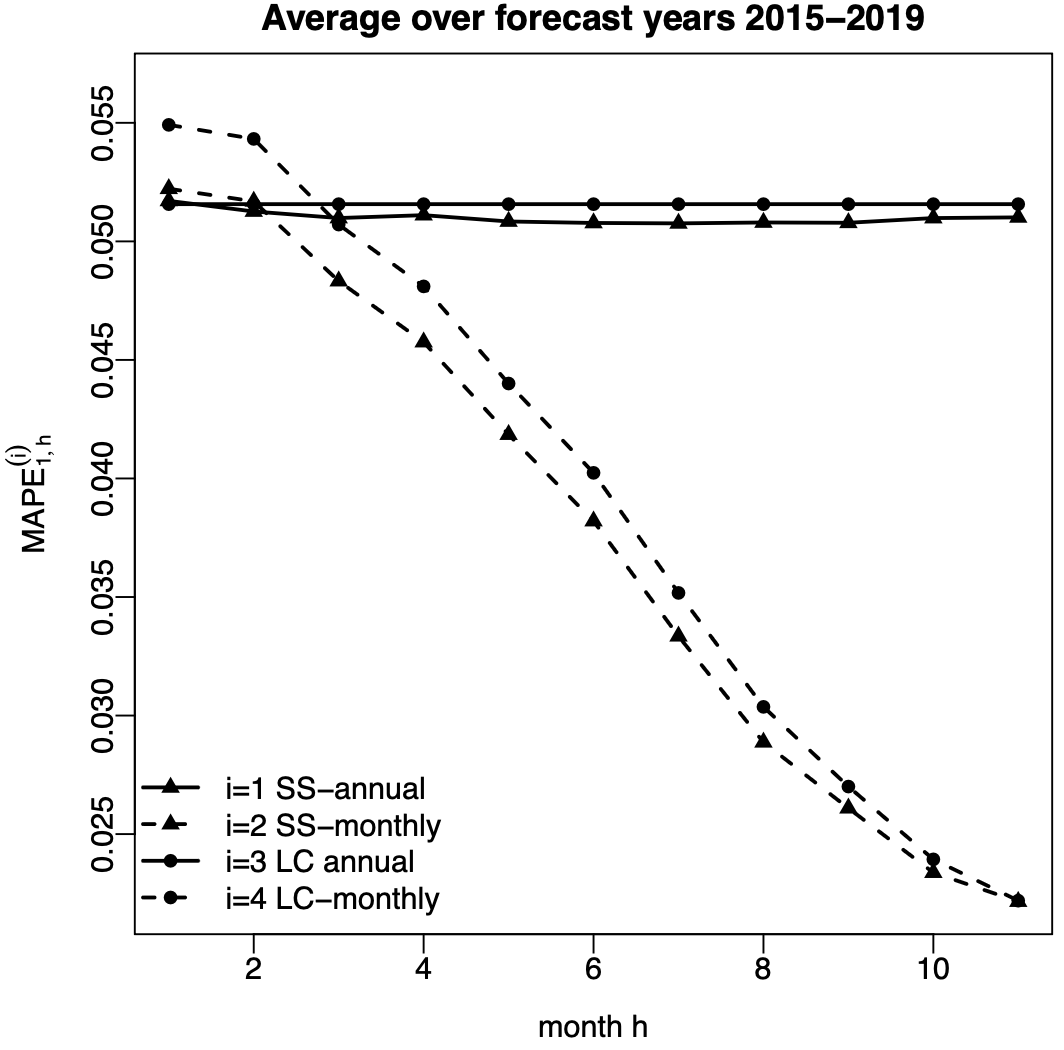}
    
\end{figure}

The left panel of Figure ~\ref{fig:forecast_one_step} plots the MAPEs for the 2015 nowcast, $\text{MAPE}_{1|(2015,h)}^{(i)}$, using the four approaches when the monthly data are observed up to month $h$ of 2015. The top half of Table~\ref{tab:monthly_horizon_results} reports the corresponding numerical values.
LC--annual is flat in $h$, as expected. MAPEs from LC--monthly decrease quickly as $h$ increases, reflecting that an increasing fraction of the year is observed and only a shrinking remainder must be forecast. In the first few months, MAPEs from LC--monthly is comparable to and can be slightly worse than those from LC-annual, but it becomes clearly more accurate later in the year.

The SS--annual forecast achieves lower MAPE than LC--annual throughout. However, its MAPEs change only modestly with $h$ and do not exhibit the strong monotone decline seen in the SS-monthly forecast. This is consistent with SS--annual updating the annual nowcast through the latent-state structure, rather than directly updating the realised deaths of the year in the annual total. By contrast, SS--monthly improves steadily over $h$ and, after the early months, it is clearly more accurate than SS--annual. 

Most importantly for assessing the value of combining annual and monthly information, SS--monthly lies below LC--monthly, with the advantage being most pronounced at the beginning of the year. This comparison shed light on the contribution of utilising annual information. Both SS--monthly and LC--monthly directly incorporate realised deaths of the current year, but SS--monthly generates the remaining-month projections from a joint model that is informed by the annual series. When $h$ is small and little current-year information is available, borrowing strength from the annual component provides a stabilising signal for the remaining-month projections, resulting in lower MAPE. As $h$ approaches year-end, the two monthly approaches converge to very similar accuracy because the annual nowcast becomes increasingly dominated by observed months.

To assess whether the 2015 pattern is representative, we repeat the same nowcasting exercise using the expanding-window design with forecast target years of $\{2015,\ldots,2019\}$. 
The right panel of Figure~\ref{fig:forecast_one_step} reports  the corresponding average MAPE across forecast target years, $\mathrm{MAPE}^{(i)}_{1,h}$, with each point summarising the accuracy of nowcast made at the end of month $h$. The bottom half of Table~\ref{tab:monthly_horizon_results} reports the corresponding numerical values.
These results reinforce four key messages. First, LC--annual is constant in $h$, reflecting the absence of intra-year updating. Second, both LC--monthly and SS--monthly show a strong decline in MAPE as $h$ increases because forecasts improve steadily as more of the current year becomes known and fewer months remain to be predicted. Third, SS--monthly is consistently  more accurate than LC--monthly, and this advantage is clearest at the beginning of the year when current-year monthly information is sparse. This highlights the benefit of exploiting both annual and monthly information when current-year monthly data are sparse. As $h$ increases, the gap narrows and the two monthly approaches converge near year-end as the annual nowcast becomes dominated by realised months. Finally, SS--annual is also relatively stable across $h$. While it does respond to intra-year monthly observations via the latent-state update, the effect on annual forecast accuracy is much smaller than the improvement obtained by anchoring to realised monthly deaths.



\begin{table}[!htbp]
\centering
\caption{Nowcast MAPE (in \%) for forecast year 2015 and the average nowcast MAPE across five expanding windows with forecast years 2015–2019. For each forecast-origin month $h$, the lowest MAPE is shaded in gray.}
\label{tab:monthly_horizon_results}
\small
\begin{tabular}{lccccccccccc}
\toprule
 & \multicolumn{11}{c}{Month $h$} \\
\cmidrule(lr){2-12}
& $h=1$ & $h=2$ & $h=3$ & $h=4$ & $h=5$ & $h=6$ & $h=7$ & $h=8$ & $h=9$ & $h=10$ & $h=11$ \\
\midrule
\multicolumn{12}{l}{$ \mathrm{MAPE}^{(i)}_{1 \mid (2015,h)}$ (\%)} \\
\midrule
SS-annual   & 3.738 & \cellcolor{gray!20}3.623 & \cellcolor{gray!20}3.539 & \cellcolor{gray!20}3.518 & 3.509 & 3.505 & 3.513 & 3.535 & 3.554 & 3.595 & 3.561 \\
SS-monthly  & \cellcolor{gray!20}3.571 & 3.770 & 3.685 & 3.649 & \cellcolor{gray!20}3.357 & \cellcolor{gray!20}3.187 & \cellcolor{gray!20}2.913 & \cellcolor{gray!20}2.609 & \cellcolor{gray!20}2.436 & \cellcolor{gray!20}2.349 & \cellcolor{gray!20}2.200 \\
LC-annual   & 3.942 & 3.942 & 3.942 & 3.942 & 3.942 & 3.942 & 3.942 & 3.942 & 3.942 & 3.942 & 3.942 \\
LC-monthly  & 3.991 & 4.148 & 4.012 & 3.941 & 3.660 & 3.462 & 3.167 & 2.820 & 2.586 & 2.444 & 2.209 \\
\midrule
\multicolumn{12}{l}{$ \mathrm{MAPE}^{(i)}_{1,h}$ (\%)} \\
\midrule
SS-annual   & 5.171 & \cellcolor{gray!20}5.126 & 5.098 & 5.111 & 5.084 & 5.078 & 5.076 & 5.080 & 5.078 & 5.098 & 5.101 \\
SS-monthly  & 5.222 & 5.169 & \cellcolor{gray!20}4.832 & \cellcolor{gray!20}4.575 & \cellcolor{gray!20}4.185 & \cellcolor{gray!20}3.820 & \cellcolor{gray!20}3.335 & \cellcolor{gray!20}2.887 & \cellcolor{gray!20}2.609 & \cellcolor{gray!20}2.337 & \cellcolor{gray!20}2.215 \\
LC-annual   & \cellcolor{gray!20}5.157 & 5.157 & 5.157 & 5.157 & 5.157 & 5.157 & 5.157 & 5.157 & 5.157 & 5.157 & 5.157 \\
LC-monthly  & 5.491 & 5.432 & 5.071 & 4.810 & 4.401 & 4.024 & 3.517 & 3.037 & 2.700 & 2.393 & 2.219 \\
\bottomrule
\end{tabular}
\end{table}

\subsubsection{Multi-year-ahead forecasts performance} 
We next examine whether the benefits of intra-year updating and utilising information from both frequencies persist for longer horizons. Specifically, we consider multi-year-ahead forecasts with horizons $n\in\{1,2,3,4,5\}$ under an expanding-window design, and we evaluate forecast accuracy for target years $\mathcal{T}\in\{2015,\ldots,2019\}$.

For each forecast origin year $t=\mathcal{T}-n+1$ and each chosen update month $h\in\{2,6,10\}$, representing early-, mid-, and late-year information sets, models are estimated using all data available up to the end of year $t-1$. At month $h$ of year $t$, the models that utilise monthly information (SS--monthly, SS--annual, and LC--monthly) update monthly factors using observations available up to the month, and we then produce forecasts for the annual mortality rates in the future target year $t-1+n$. 
This setup is similar to the nowcast exercise in the sense that forecasts are updated as additional monthly information arrives during the forecast origin year. The key difference is that, for $n\ge 2$, none of the target-year months have been observed at the forecast origin. Consequently, the forecasts from SS--monthly and LC--monthly are constructed by aggregating monthly forecasts for the future target year, rather than combining realised and forecasted values, as in the nowcast case. Even so, the update at month $h$ remains relevant. Conditioning on more months in year $t$ refines the current monthly factor value, which then affects the projected paths into $t-1+n$.

\begin{figure}[!htbp]
    \centering
    \caption{Average $n$-year-ahead forecast MAPE across expanding windows with forecast years of 2015-2019, evaluated at forecast origin months $h=2,6,10$.}
    \label{fig:multi_step_rolling}
    \includegraphics[width=0.49\textwidth]{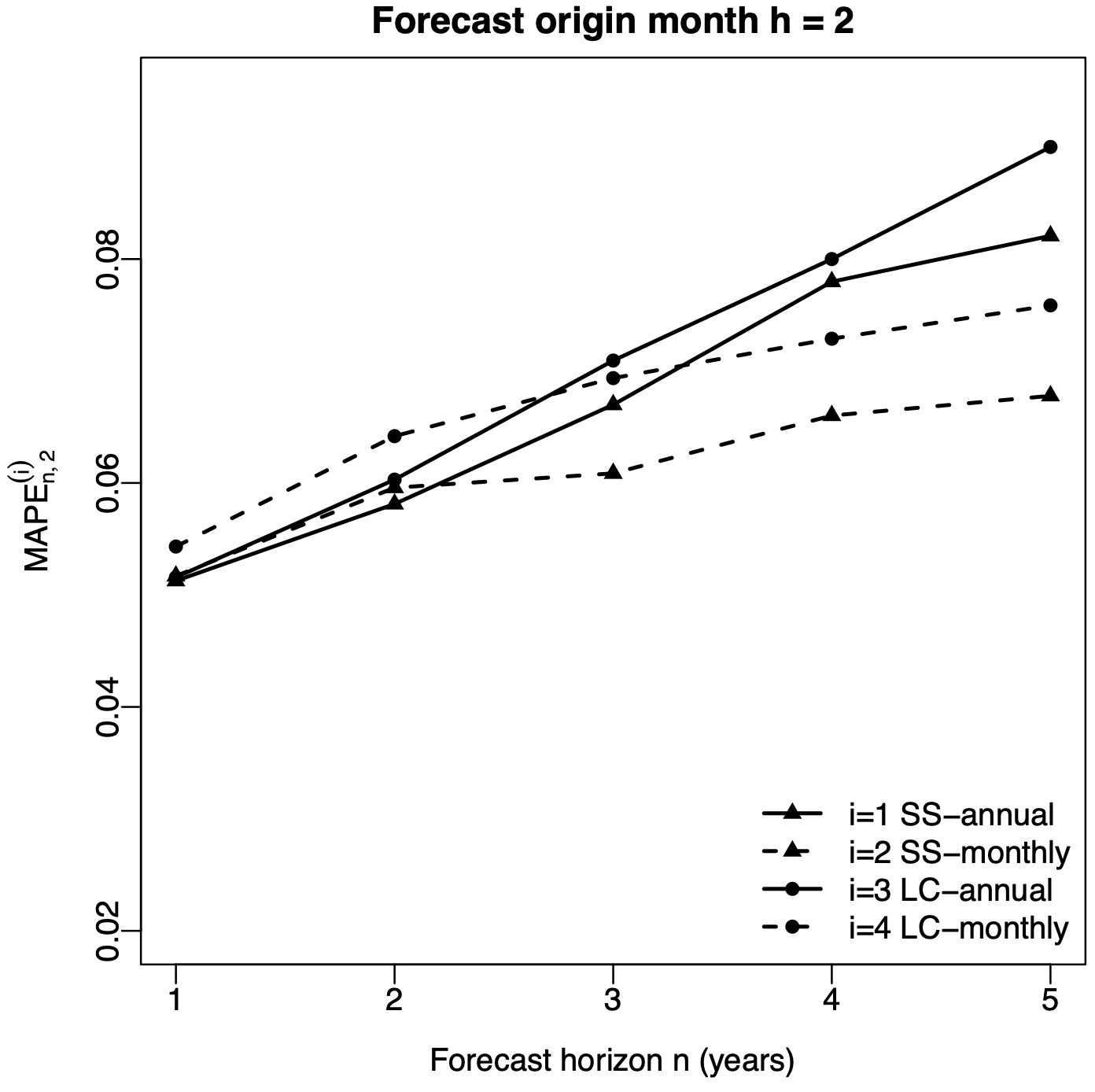}
    \includegraphics[width=0.49\textwidth]{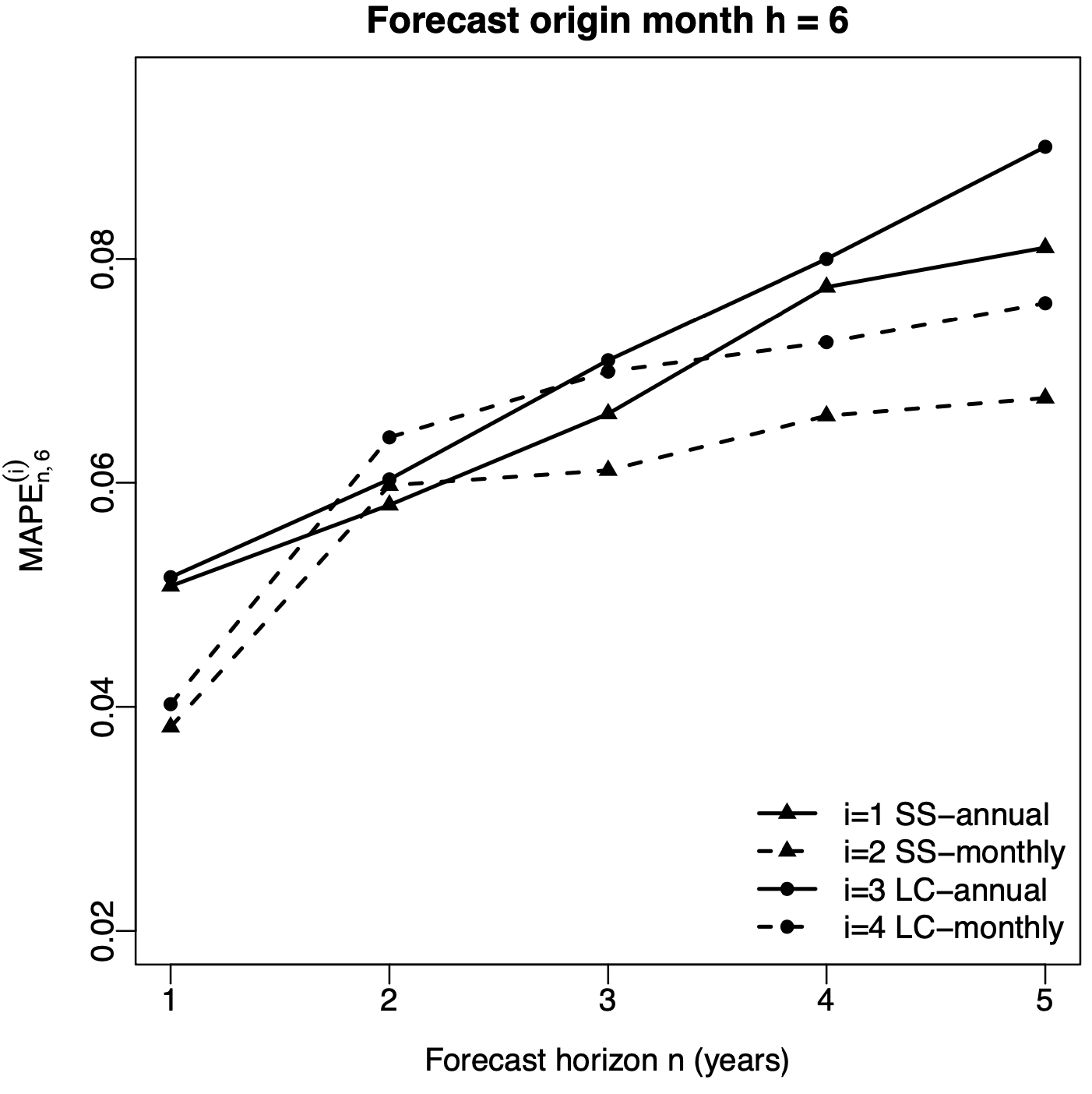}
    \includegraphics[width=0.49\textwidth]{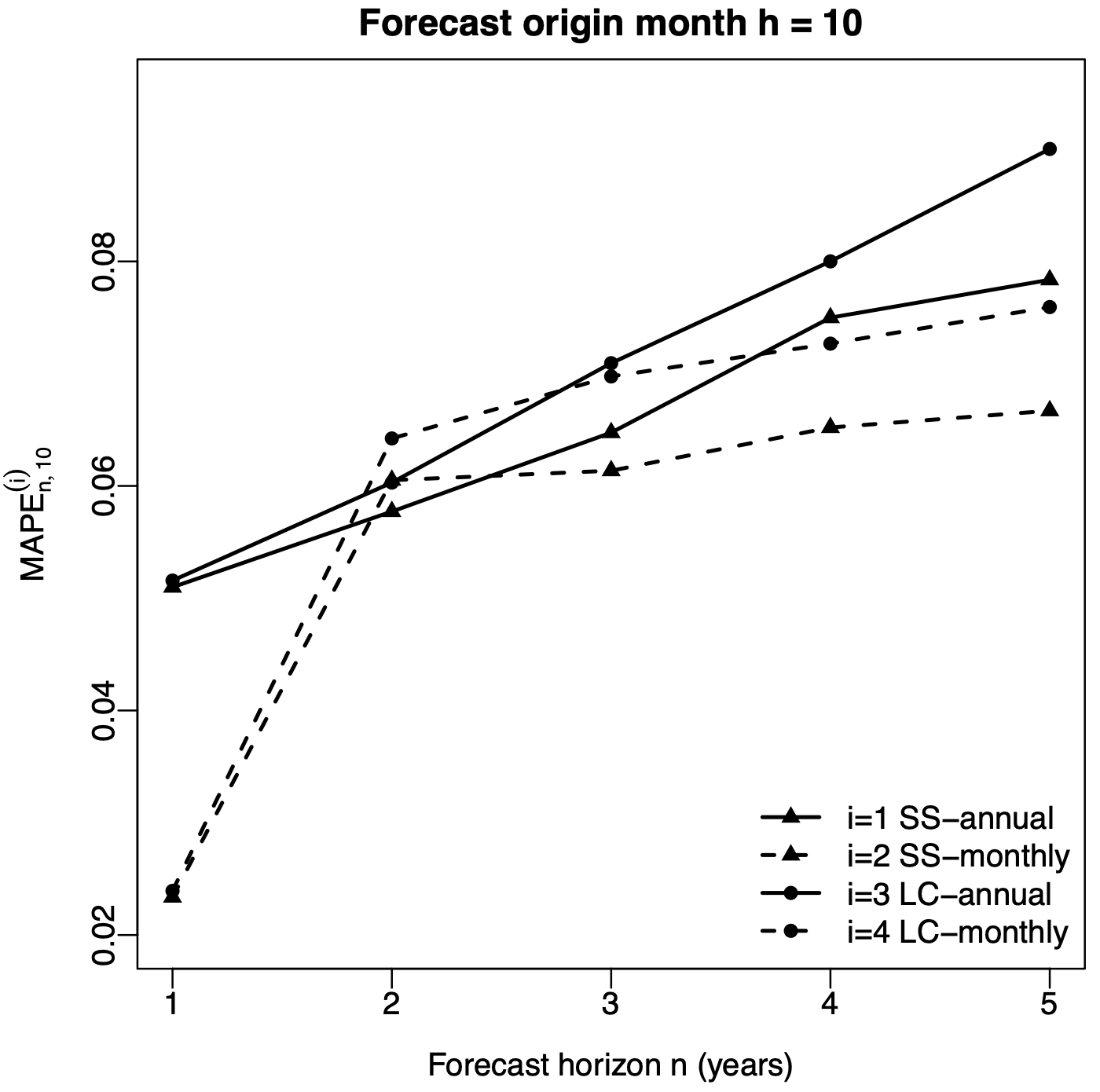}
\end{figure}

\begin{table}[!htbp]
\centering
\caption{Average $n$-year-ahead forecast MAPE (in \%) across expanding windows with forecast years of 2015-2019.  For each forecast horizon $n$, the lowest MAPE is shaded in gray.}
\label{tab:mape_month_comparison}
\small
\begin{tabular}{lccccc}
\toprule
 & \multicolumn{5}{c}{Forecast horizon $n$ (years)} \\
\cmidrule(lr){2-6}
MAPE (\%) & $n=1$ & $n=2$ & $n=3$ & $n=4$ & $n=5$ \\
\midrule
\multicolumn{6}{l}{MAPE$^{(i)}_{n,2}$ (forecast origin month 2)} \\
\midrule
SS-annual   & \cellcolor{gray!20}5.126 & \cellcolor{gray!20}5.812 & 6.699 & 7.798 & 8.208 \\
SS-monthly  & 5.169 & 5.958 & \cellcolor{gray!20}6.086 & \cellcolor{gray!20}6.602 & \cellcolor{gray!20}6.778 \\
LC-annual   & 5.157 & 6.030 & 7.093 & 8.001 & 9.002 \\
LC-monthly  & 5.432 & 6.418 & 6.936 & 7.288 & 7.586 \\
\midrule
\multicolumn{6}{l}{MAPE$^{(i)}_{n,6}$ (forecast origin month 6)} \\
\midrule
SS-annual   & 5.078 & \cellcolor{gray!20}5.801 & 6.616 & 7.749 & 8.102 \\
SS-monthly  & \cellcolor{gray!20}3.820 & 5.977 & \cellcolor{gray!20}6.111 & \cellcolor{gray!20}6.597 & \cellcolor{gray!20}6.756 \\
LC-annual   & 5.157 & 6.030 & 7.093 & 8.001 & 9.002 \\
LC-monthly  & 4.024 & 6.405 & 6.992 & 7.255 & 7.605 \\
\midrule
\multicolumn{6}{l}{MAPE$^{(i)}_{n,10}$ (forecast origin month 10)} \\
\midrule
SS-annual   & 5.098 & \cellcolor{gray!20}5.772 & 6.475 & 7.499 & 7.836 \\
SS-monthly  & \cellcolor{gray!20}2.337 & 6.051 & \cellcolor{gray!20}6.136 & \cellcolor{gray!20}6.520 & \cellcolor{gray!20}6.670 \\
LC-annual   & 5.157 & 6.030 & 7.093 & 8.001 & 9.002 \\
LC-monthly  & 2.393 & 6.423 & 6.975 & 7.267 & 7.594 \\
\bottomrule
\end{tabular}
\end{table}

Figure~\ref{fig:multi_step_rolling} reports the average multi-year-ahead forecast MAPE across expending windows, evaluated at three forecast-origin months ($h=2,6,10$). In each panel, the horizontal axis is the forecast horizon $n$ (in years) and the vertical axis is $\mathrm{MAPE}_{n,h}$, averaged across expanding windows. For example, the point at $(n=3,h=2)$ summarises the mean MAPE of three-year-ahead forecasts produced using information available up to month 2 for forecast-origin years 2013, 2014, $\ldots$, 2017, and therefore targets years 2015, 2016, $\ldots$, 2019. Table~\ref{tab:mape_month_comparison} reports the corresponding numerical values.

Several patterns emerge. First, across all three panels and for all methods, forecasting errors $\mathrm{MAPE}_{n,h}^{(i)}$ increase with the forecast horizon $n$. This deterioration is expected, since uncertainty accumulates as forecasts are pushed further into the future. 

Second, the monthly-based approaches experience a sharp improvement at $n=1$ as $h$ moves from 2 to 6 to 10 due to incorporating more realised months into nowcast and forecasting for fewer months.
In contrast, the annual benchmarks are comparatively insensitive to $h$ at $n=1$. This reflects that they do not benefit from the direct anchoring of the annual total to realised intra-year months.

Third, the value of utilising monthly data in the forecast origin year is concentrated in the $n=1$ nowcast. In the month-$6$ and month-$10$ panels, the monthly-based curves show a noticeable increase in MAPE from $n=1$ to $n=2$. At $n=2$ and beyond, the target year is entirely in the future, so none of its months are observed at the forecast origin and the monthly-based approaches no longer benefit from substituting realised months into the annual total. The corresponding increase is much smaller in the month-$2$ panel because the $n=1$ nowcast is only weakly anchored when just two months have been observed.

Finally, utilising both annual and monthly information improves performance relative to using monthly information alone. In all three panels, SS--monthly always achieves lower MAPE than LC--monthly, and the advantage is more visible at the beginning of the year ($h=2$), when current-year monthly information is sparse and a monthly-only model has less signal to stabilise its projected path.

Overall, Figure~\ref{fig:multi_step_rolling} complements the one-step nowcast analysis. Intra-year information provides the largest gains for $n=1$ through direct anchoring, while for genuinely multi-year forecasting ($n\ge 2$) performance differences are driven more by model structure. The SS-monthly approach that exploits both annual and monthly information typically achieves the lowest or second lowest forecast error.

\section{Reconciling annual and monthly forecasts}
\label{sec:reconciliation}

\subsection{Forecast reconciliation for temporal hierarchy}

Forecast reconciliation adjusts a set of forecasts so that they respect the aggregation relationships implied by a hierarchy. In this study, the hierarchy is temporal: annual total deaths are the aggregate of monthly deaths within the same calendar year. For each age $x$ and year $t$, the summing constraint is
\[
D_{x,t}=\sum_{h=1}^{12} d_{x,(t,h)}.
\]
Reconciled forecasts enforce this summing constraint at each forecast horizon, so that the annual forecast equals the sum of the corresponding monthly forecasts.

Earlier work in forecast reconciliation is achieved by primarily forecasting one level of the hierarchy, which was then used to produce projections at other levels. \cite{GG1960} advocate the top-down approach, where model forecasts are performed only for the most aggregated series, which is then disaggregated by historical proportion to produce projections at lower levels. On the other hand, bottom-up approach, preferred by \cite{Dunn1976} and \cite{Zellner2000}, forecasts the most disaggregated time series and sums them up as predictions at higher levels of the hierarchy. \cite{DASILVA2019} highlight the better interpretation of the bottom-up approach as an additional advantage over the top-down approach.

A key development is the optimal combination approach of \cite{hyndman2011}, which allows independent base forecasts to be produced at all levels and then reconciled by optimally combining these forecasts subject to the aggregation constraints. The reconciliation weights depend on the hierarchical structure and the relative uncertainty of the base forecasts. \cite{hyndman2011} demonstrate improved performance relative to purely top-down or bottom-up strategies in many settings.

While most of the reconciliation literature is framed for cross-sectional hierarchies. \cite{ATHANASOPOULOS201760} extend reconciliation strategies to temporal hierarchies. They treat temporal aggregation (e.g., annual--monthly) as a hierarchy governed by a summing matrix, and show how reconciliation methods such as optimal combination can be applied  across temporal levels. We follow this temporal-reconciliation framework to reconcile annual and monthly death forecasts obtained from annual and monthly forecasts.

In what follows, we suppress the age index $x$ for notational simplicity and reconciliation is performed separately for each age. Let $\mathbf{D}_t$ collect all series in the temporal hierarchy (annual and monthly) for year $t$, and let $\mathbf{d}_t$ denote the vector of bottom-level (monthly) series:
\[
\mathbf{D}_t=\bigl[D_t,\ d_{(t,1)},\ldots,d_{(t,12)}\bigr]^\top,
\qquad
\mathbf{d}_t=\bigl[d_{(t,1)},\ldots,d_{(t,12)}\bigr]^\top.
\]
The temporal hierarchy can be written compactly as
\begin{equation}
\mathbf{D}_t=\mathbf{S}\mathbf{d}_t,
\label{eq:coherence}
\end{equation}
where $\mathbf{S}$ is the summing matrix mapping the bottom-level series into all levels:
\[
\mathbf{S}\in\mathbb{R}^{13\times 12},\qquad
\mathbf{S}=
\begin{bmatrix}
\mathbf{1}_{12}^\top\\
\mathbf{I}_{12}
\end{bmatrix}.
\]
Here, $\mathbf{1}_{12}=(1,\ldots,1)^\top$, and $\mathbf{I}_{12}$ denotes the $12\times12$ identity matrix. 

The bottom-up approach forecasts all bottom-level series and then aggregates them to obtain forecasts at higher levels \citep{Dunn1976,Zellner2000}. Let $\hat{d}_{(t+n,h)\,|\,t}$ be the $n$-year-ahead forecast for month $h$ in year $t+n$, formed at the end of year $t$. The corresponding bottom-up forecast of annual deaths in year $t+n$ is
\begin{equation}
\hat{D}^{\text{BU}}_{t+n\,|\,t}=\sum_{h=1}^{12}\hat{d}_{(t+n,h)\,|\,t}.
\label{eq:BU}
\end{equation}

The aggregated monthly forecast used in this study can be viewed as a bottom-up construction, since it first builds annual total deaths from monthly death forecasts and then convert them to annual mortality rates. 

\subsection{Optimal combination reconciliation}
\label{sec:opti_comb}
Different from the bottom-up aggregation, the optimal combination method begins from base forecasts produced independently at multiple levels and then reconciles them in a statistically principled way.

Following the optimal combination framework of \cite{hyndman2011} and its temporal-hierarchy implementation in \cite{ATHANASOPOULOS201760}, let $\hat{\mathbf{D}}_{t+k\,|\,t}$ denote the vector of base forecasts produced independently at each level. The approach assumes the regression representation
\begin{equation}
\hat{\mathbf{D}}_{t+k\,|\,t}=\mathbf{S}\boldsymbol{\beta}_{t+k\,|\,t}+\boldsymbol{\varepsilon}_{t+k\,|\,t},
\qquad
\mathbb{E}(\boldsymbol{\varepsilon}_{t+k\,|\,t})=\mathbf{0},
\qquad
\mathrm{Var}(\boldsymbol{\varepsilon}_{t+k\,|\,t})=\mathbf{W}_{k},
\label{eq:recon_reg}
\end{equation}
where $\boldsymbol{\beta}_{t+k\,|\,t}\in\mathbb{R}^{12}$ is the unknown conditional mean of the future values of the bottom-level observed series and $\mathbf{W}_{k}$ is the covariance matrix of base forecast errors at horizon $k$.

Reconciled forecasts are obtained via generalised least squares:
\begin{equation}
\tilde{\boldsymbol{\beta}}_{t+k\,|\,t}
=
\arg\min_{\boldsymbol{\beta}}
\left(\hat{\mathbf{D}}_{t+k\,|\,t}-\mathbf{S}\boldsymbol{\beta}\right)^{\top}
\mathbf{W}_{k}^{-1}
\left(\hat{\mathbf{D}}_{t+k\,|\,t}-\mathbf{S}\boldsymbol{\beta}\right),
\end{equation}
which yields the reconciled vector
\begin{equation}
\tilde{\mathbf{D}}_{t+k\,|\,t}
=
\mathbf{S}\tilde{\boldsymbol{\beta}}_{t+k\,|\,t}
=
\mathbf{S}\left(\mathbf{S}^{\top}\mathbf{W}_{k}^{-1}\mathbf{S}\right)^{-1}\mathbf{S}^{\top}\mathbf{W}_{k}^{-1}\hat{\mathbf{D}}_{t+k\,|\,t}.
\label{eq:opt-comb-general}
\end{equation}

Equation \eqref{eq:opt-comb-general} produces forecasts that satisfy the aggregation constraints exactly, while combining information across the annual and monthly base forecasts according to their relative uncertainty through $\mathbf{W}_{k}$.

Implementing \eqref{eq:opt-comb-general} requires an estimate of the covariance matrix of base forecast errors. Following the temporal reconciliation approach of \cite{ATHANASOPOULOS201760}, we estimate a single weight matrix $\mathbf{W}$ from in-sample one-step-ahead base forecast errors and use this estimate for all forecast horizons considered.

For each age, we first construct base forecasts at the annual and monthly levels using information available at the end of the previous calendar year. For example, the annual Lee--Carter model produces an annual base forecast for year $t$,
$\hat{D}^{(A)}_{t\,|\,t-1}$,
formed using annual data up to year $t-1$. Independently, the monthly Lee--Carter model produces base forecasts for each month $h=1,\ldots,12$ of year $t$,
$
\hat{d}^{(M)}_{(t,h)\,|\,t-1},
$
formed using monthly data available up to year $t-1$. Thus, all monthly forecasts for year $t$ are generated without conditioning on any intra-year observations from year $t$.

We collect the annual and monthly base forecasts into the vector
\[
\hat{\mathbf{D}}_{t\,|\,t-1}
=
\bigl[
\hat{D}^{(A)}_{t\,|\,t-1},\
\hat{d}^{(M)}_{(t,1)\,|\,t-1},\ldots,\hat{d}^{(M)}_{(t,12)\,|\,t-1}
\bigr]^\top.
\]

Define the vector of base forecast errors
\[
\mathbf{e}_t=
\bigl[
e^{(A)}_t,\ e^{(M)}_{(t,1)},\ldots,e^{(M)}_{(t,12)}
\bigr]^\top,
\qquad t=1,\ldots,T,
\]
where
\[
e^{(A)}_{t}=D_{t}-\hat{D}^{(A)}_{t\,|\,t-1},
\qquad
e^{(M)}_{(t,h)}=d_{(t,h)}-\hat{d}^{(M)}_{(t,h)\,|\,t-1}.
\]
A natural estimator of the covariance is the sample covariance matrix
\begin{equation}
\hat{\mathbf{W}}
=
\frac{1}{t_1-t_0}\sum_{t=t_0+1}^{t_1}\mathbf{e}_t\mathbf{e}_t^\top.
\label{eq:W_full}
\end{equation}
Because $\hat{\mathbf{W}}$ is $13\times13$ and may be imprecisely estimated in short samples, we adopt the series variance scaling approach of \cite{ATHANASOPOULOS201760}. This approach assumes $\mathbf{W}$ is diagonal and that base forecast errors share a common variance within each aggregation level. Accordingly, we set
\begin{equation}
\hat{\mathbf{W}}
=
\mathrm{diag}\!\Bigl(
(\hat{\sigma}^{(A)})^2,\ (\hat{\sigma}^{(M)})^2\mathbf{1}_{12}
\Bigr),
\label{eq:W_scaled}
\end{equation}
where $(\hat{\sigma}^{(A)})^2$ is estimated from $\{e^{(A)}_t\}_{t=t_0+1}^{t_1}$ and $(\hat{\sigma}^{(M)})^2$ is estimated from $\{e^{(M)}_{(t,h)}\}_{t=t_0+1,\ldots,t_1;\ h=1,\ldots,12}$.

\subsection{Forecast reconciliation performance}
As noted earlier, aggregating monthly death forecasts to obtain annual deaths (and the corresponding annual mortality rates) corresponds to the bottom-up method. In this section, unless stated otherwise, reconciliation refers to temporal reconciliation via the optimal-combination method. 

We use the same expanding-window design as in Section~\ref{sec:eval_forecast} to evaluate the benefit of temporal reconciliation. After producing base forecasts of annual and monthly deaths from the SS or LC model, we reconcile these forecasts to satisfy the aggregation constraint that annual deaths equal the sum of monthly deaths. Finally, we convert the reconciled annual death forecasts to annual mortality rates using the same exposure construction as in Section~\ref{sec:forecast_type}.

\begin{figure}[!htbp]
\centering
\caption{Average $n$-year-ahead forecast  MAPE  across expanding windows with forecast years of 2015-2019, for direct annual, aggregated monthly, and temporally reconciled forecasts, evaluated at forecast origin months $h=2,6,10$. }
\label{fig:multi_step_rolling_reconciled}
\includegraphics[width=0.9\textwidth]{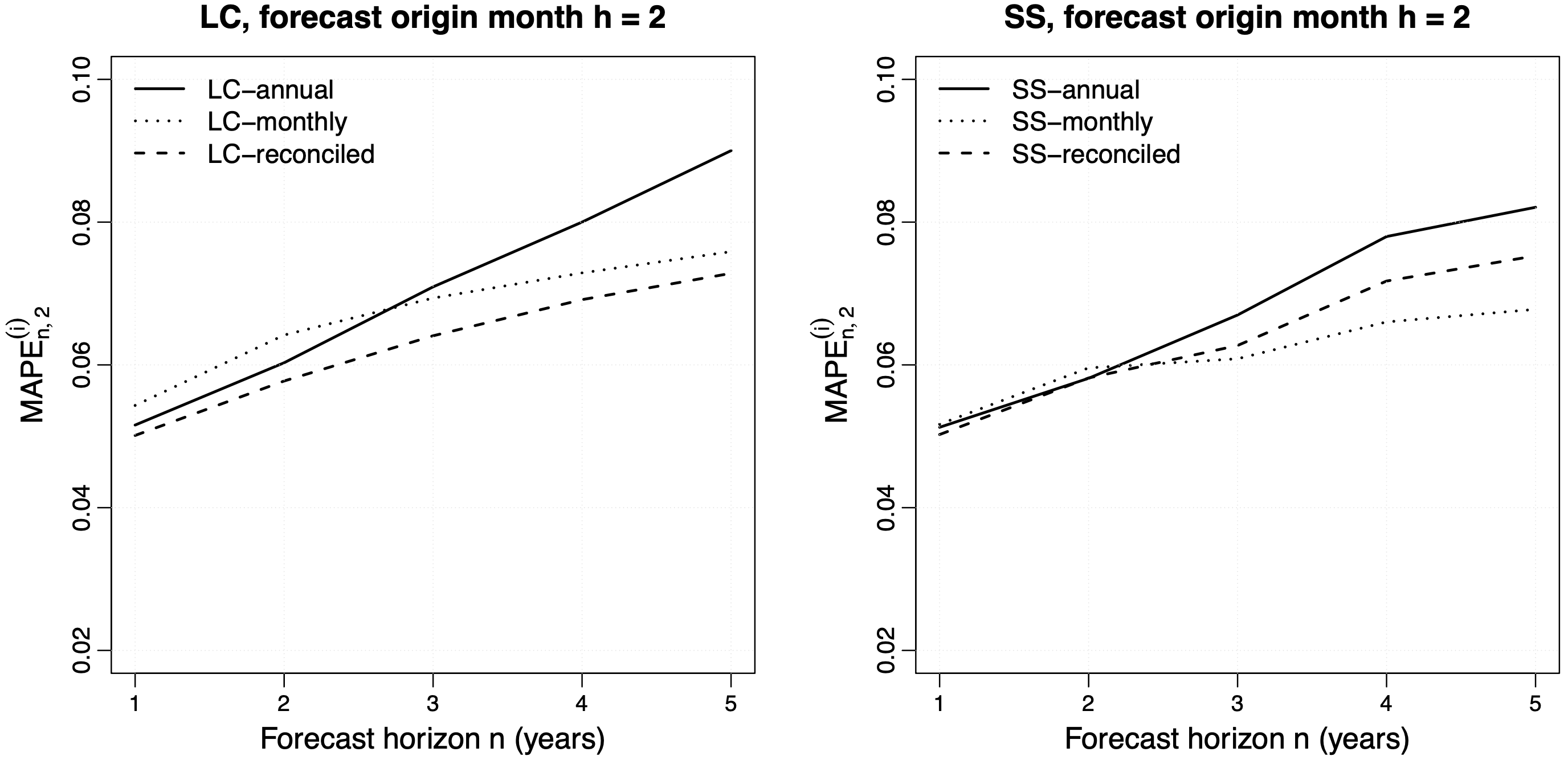}\par\medskip
\includegraphics[width=0.9\textwidth]{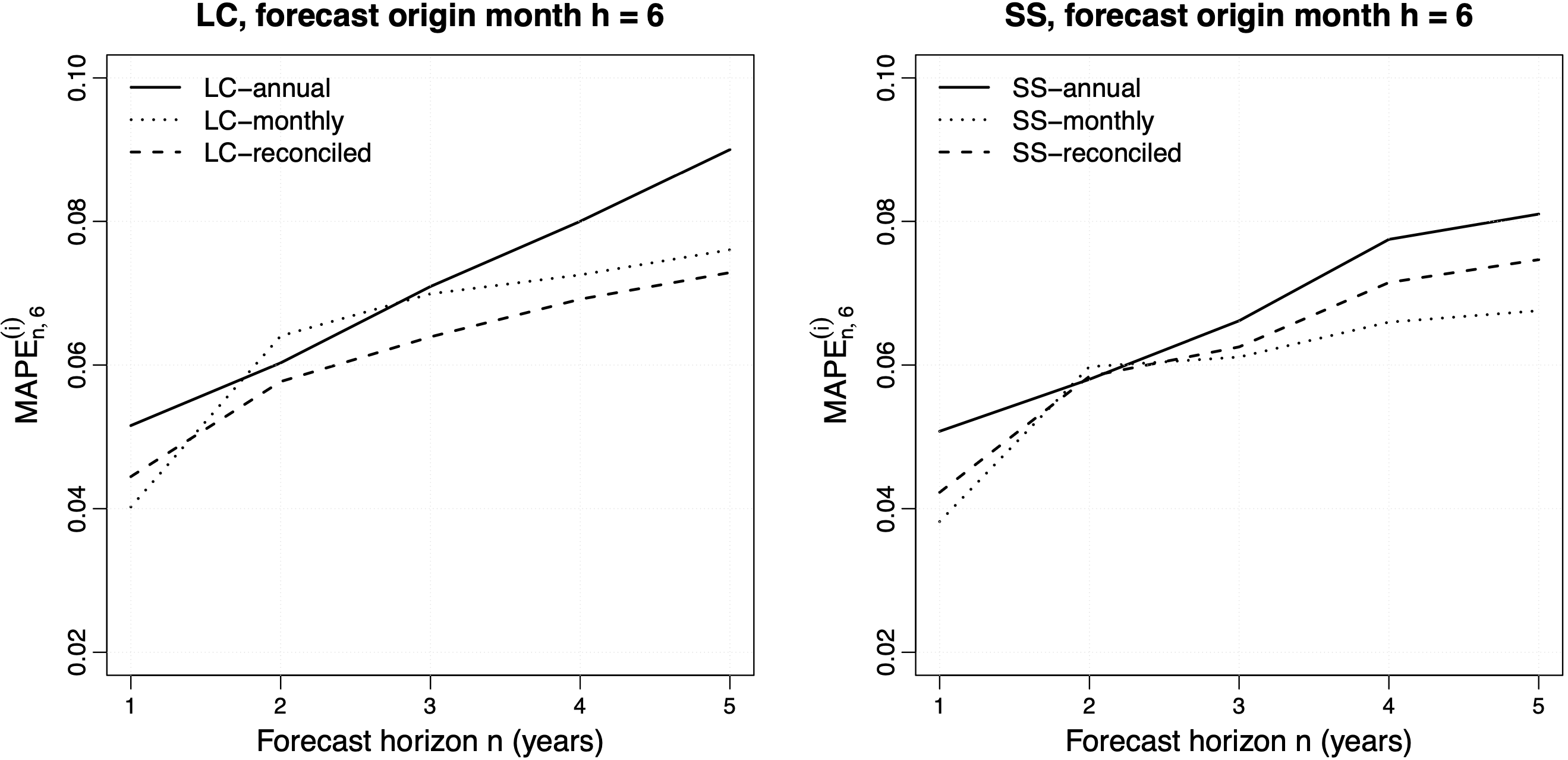}\par\medskip
\includegraphics[width=0.9\textwidth]{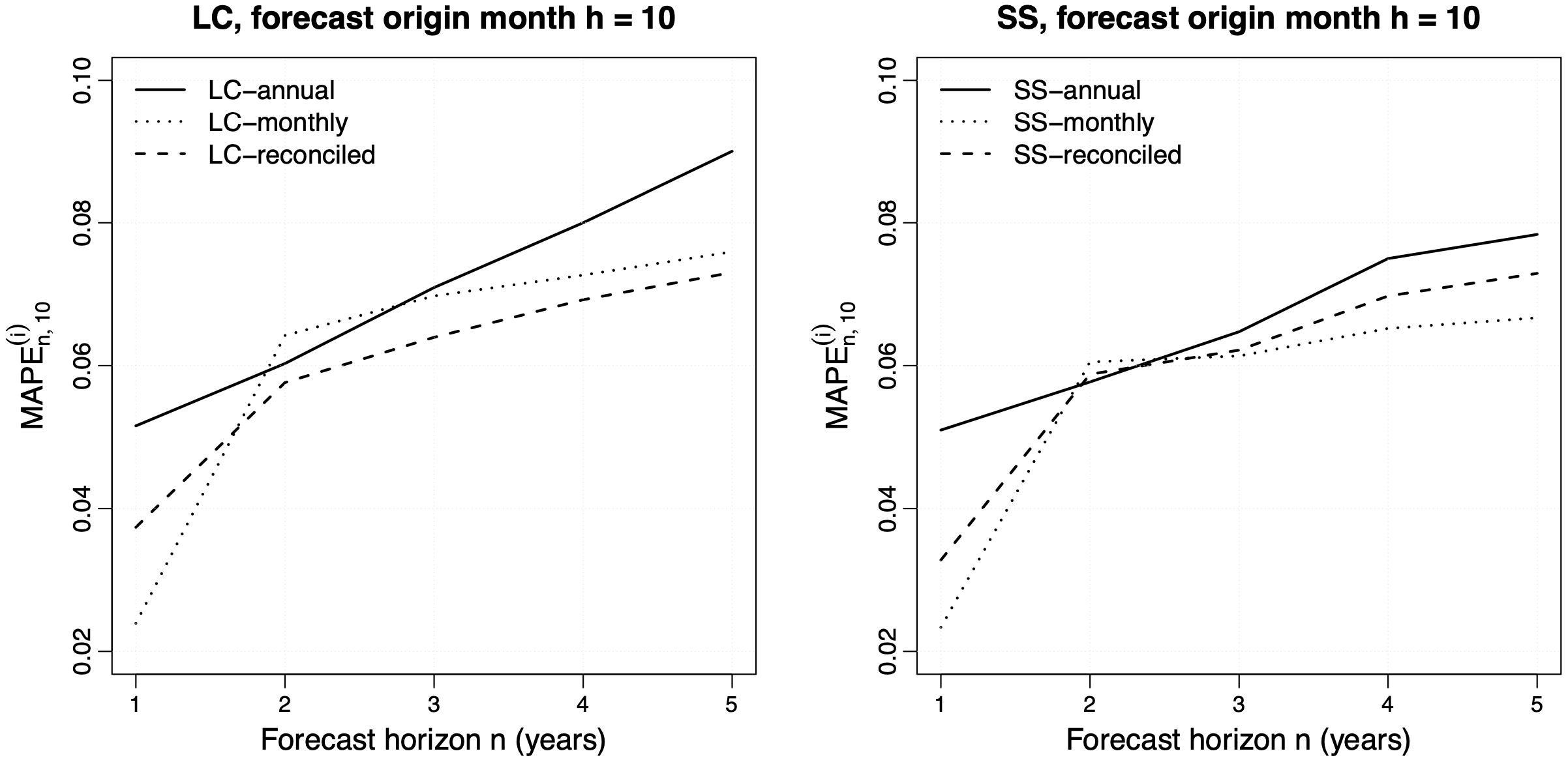}
\end{figure}

Figure~\ref{fig:multi_step_rolling_reconciled} compares the MAPE for annual mortality forecasts with or without reconciliation, averaged over five expanding windows. In each panel, the solid, dotted, and dashed lines correspond to direct annual forecasts, aggregated monthly forecasts, and temporally reconciled forecasts, respectively. The left column shows results for the independent LC models and the right column for the SS model. Rows correspond to forecast-origin months $h=2,6,10$. Compared with Figure~\ref{fig:multi_step_rolling}, this figure adds the MAPE for the reconciled forecasts and present the results by the model choice.
Table \ref{tab:mape_recon_month_comparison} presents the numerical values of the average MAPE of reconciled forecasts over the expanding windows. 

\begin{table}[!htbp]
\centering
\caption{Average reconciled $n$-year-ahead forecast MAPE (in \%) across expanding windows with forecast years of 2015-2019.}
\label{tab:mape_recon_month_comparison}
\small
\begin{tabular}{lccccc}
\toprule
 & \multicolumn{5}{c}{Forecast horizon $n$ (years)} \\
\cmidrule(lr){2-6}
MAPE (\%) & $n=1$ & $n=2$ & $n=3$ & $n=4$ & $n=5$ \\
\midrule
\multicolumn{6}{l}{MAPE$^{(i)}_{n,2}$ (forecast origin month 2)} \\
\midrule
SS-reconciled & 5.024 & 5.817 & 6.272 & 7.175 & 7.528 \\
LC-reconciled &5.012 & 5.774 & 6.408 & 6.914 & 7.283 \\
\midrule
\multicolumn{6}{l}{MAPE$^{(i)}_{n,6}$ (forecast origin month 6)} \\
\midrule
SS-reconciled &4.227 & 5.842 & 6.252 & 7.149 & 7.467 \\
LC-reconciled & 4.446 & 5.770 & 6.392 & 6.917 & 7.287 \\

\midrule
\multicolumn{6}{l}{MAPE$^{(i)}_{n,10}$ (forecast origin month 10)} \\
\midrule
SS-reconciled & 3.279 & 5.879 & 6.218 & 6.975 & 7.291 \\
LC-reconciled & 3.738 & 5.763 & 6.396 & 6.923 & 7.304 \\
\bottomrule
\end{tabular}
\end{table}

As shown in Figure~\ref{fig:multi_step_rolling_reconciled} and by the comparison of Table \ref{tab:mape_month_comparison} and Table \ref{tab:mape_recon_month_comparison}, temporal reconciliation significantly improves accuracy for the independently LC forecasts. When annual and monthly LC models are estimated separately, each level captures different aspects of the signal. By combining information from both temporal levels subject to the aggregation constraint, the reconciliation  provides a more accurate forecast.

For the SS model, reconciliation yields very limited benefit. The reconciled SS forecasts typically lie between the SS annual and SS monthly curves. Therefore, reconciliation reduces error relative to the annual SS forecasts, but it seldom improves upon the SS--monthly forecasts. This is consistent with the joint SS specification, in which annual and monthly observations are linked through a common latent process. Consequently, the two base forecasts already share information across temporal frequencies, leaving less scope for reconciliation to improve accuracy.

Finally, although the SS--monthly forecasts generally achieve lower MAPE than both of the independent LC forecasts as shown in Figure \ref{fig:multi_step_rolling}, the performance gap narrows significantly once the independent LC forecasts are reconciled. This indicates that a non-trivial part of the initial SS advantage stems from its internal cross-frequency information sharing.

\subsection{Forecast uncertainty}
We next compare prediction uncertainty for the two best-performing approaches identified in the point-forecast evaluation: the aggregated SS monthly forecasts and the temporally reconciled forecasts based on the independently estimated LC models. These methods achieved the lowest MAPE within their respective model classes. Our focus here is on their 95\% prediction intervals.

We estimate model parameters using data from January 1999 to December 2014. For a mid-year forecast origin, we update the monthly period factor using monthly observations available up to June 2015, and then generate mortality forecasts for 2015, 2016, \ldots, 2019. Prediction intervals at the 95\% level are constructed using Monte Carlo simulation with $10{,}000$ replications.

\begin{figure}[!htbp]
\centering
\caption{Point forecasts of annual mortality rates and associated 95\% prediction intervals for $x=60$ and $x=70$ in 2015-2019, based on the aggregated SS monthly forecasts and the temporally reconciled LC forecasts.}
\label{fig:prediction_interval_6ages}

\includegraphics[width=0.49\textwidth]{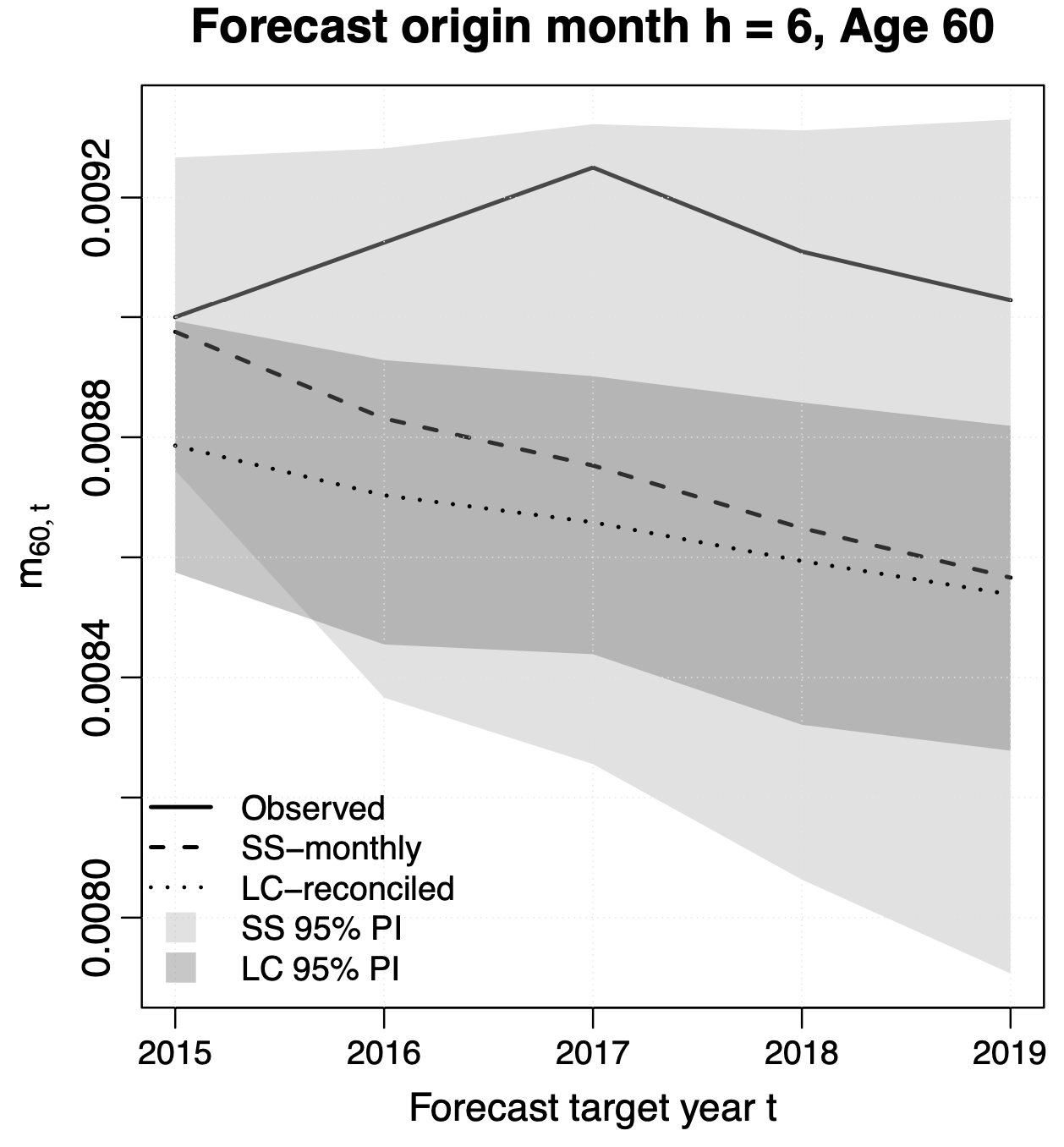}
\includegraphics[width=0.49\textwidth]{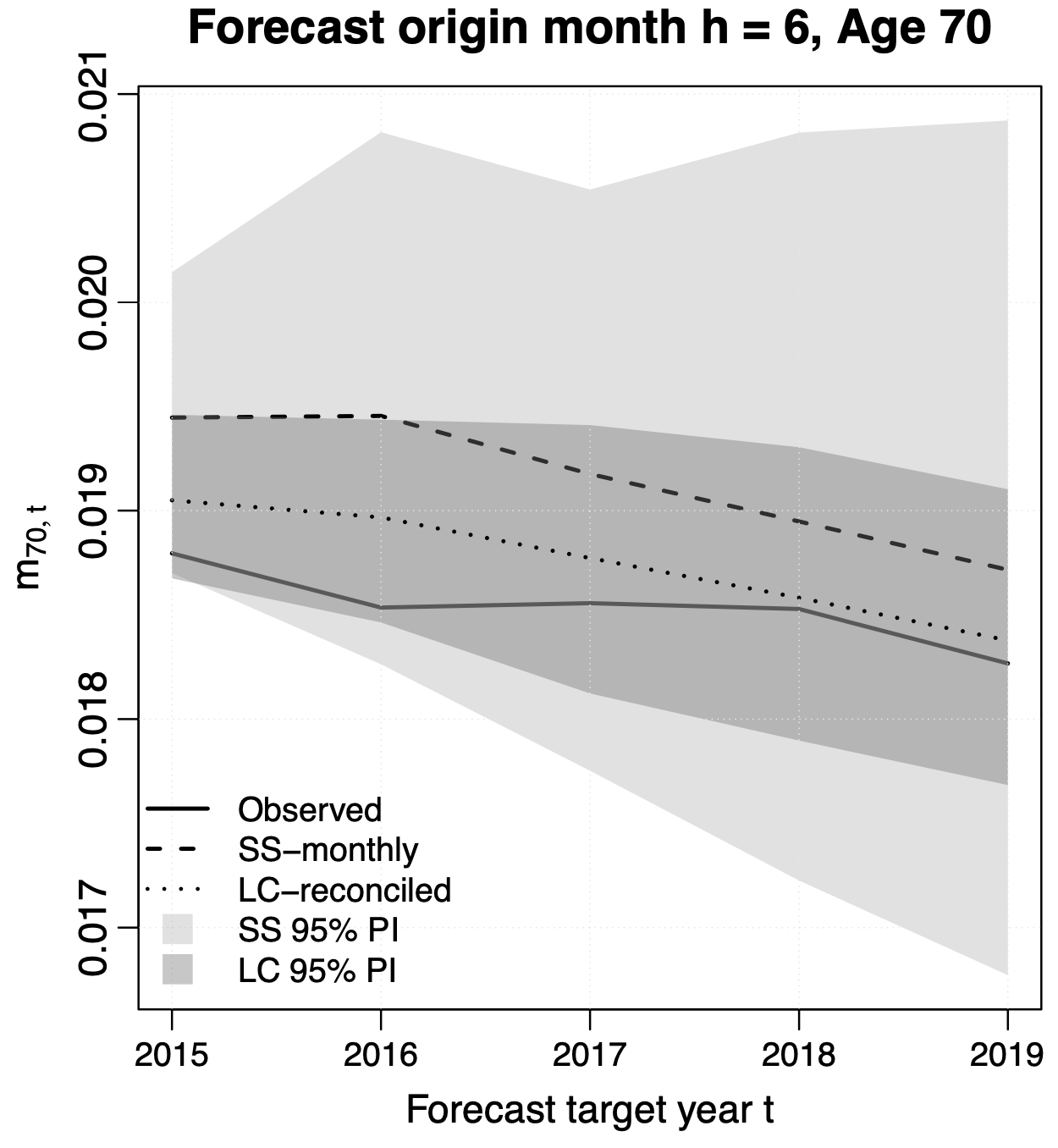}

\end{figure}

Figure~\ref{fig:prediction_interval_6ages} shows point forecasts and 95\% prediction intervals for two representative ages, $x=60$ and $x=70$, in 2015-2019 using the two approaches. The solid line denotes the realised annual mortality rate, while the dashed and dotted lines correspond to the SS--monthly and LC--reconciled point forecasts, respectively. Shaded areas indicate the associated 95\% prediction intervals.

Figure~\ref{fig:prediction_interval_6ages} shows that the SS--monthly approach produces much wider prediction intervals than the reconciled LC approach for both ages. The same pattern holds for most of the remaining ages (not shown). This difference is consistent with the estimated state dynamics. In our application, the estimated innovation variance of the monthly period factor under the SS specification is $\widehat{\sigma}^2_{\omega}=5.56$, which is much larger than the corresponding estimate from the LC--monthly model, $\widehat{\sigma}^2_{\eta_4}=1.452$. The estimated measurement-error variances in the monthly observation equation are broadly comparable across the SS and monthly LC specifications. Hence, the wider SS intervals are primarily driven by the larger state innovation variance, which accumulates over the forecast horizon.

The difference in estimated state variability is methodological. The monthly LC approach is implemented in two stages: the period factor $k_{\tau}$ is first estimated from the LC model and then treated as observed when fitting the SARIMA dynamics. Forecast uncertainty is subsequently computed conditional on this plug-in estimate. As a result, uncertainty from the first-stage extraction of $k_{\tau}$ is only partially propagated into the prediction distribution. In contrast, the SS model estimates the latent process jointly with the remaining parameters under the full likelihood. This joint fit can attribute a greater share of variability to the state innovations in order to simultaneously accommodate both annual and monthly observations, leading to a larger estimated innovation variance and, consequently, wider predictive intervals.

\begin{figure}[!htbp]
\centering
\caption{Point forecasts of annual mortality rates and associated 95\% prediction intervals for $x=60$ and $x=70$ in 2015-2019, based on the aggregated LC monthly forecasts and the temporally reconciled LC forecasts.}
\label{fig:prediction_interval_6ages_lc}
\includegraphics[width=0.49\textwidth]{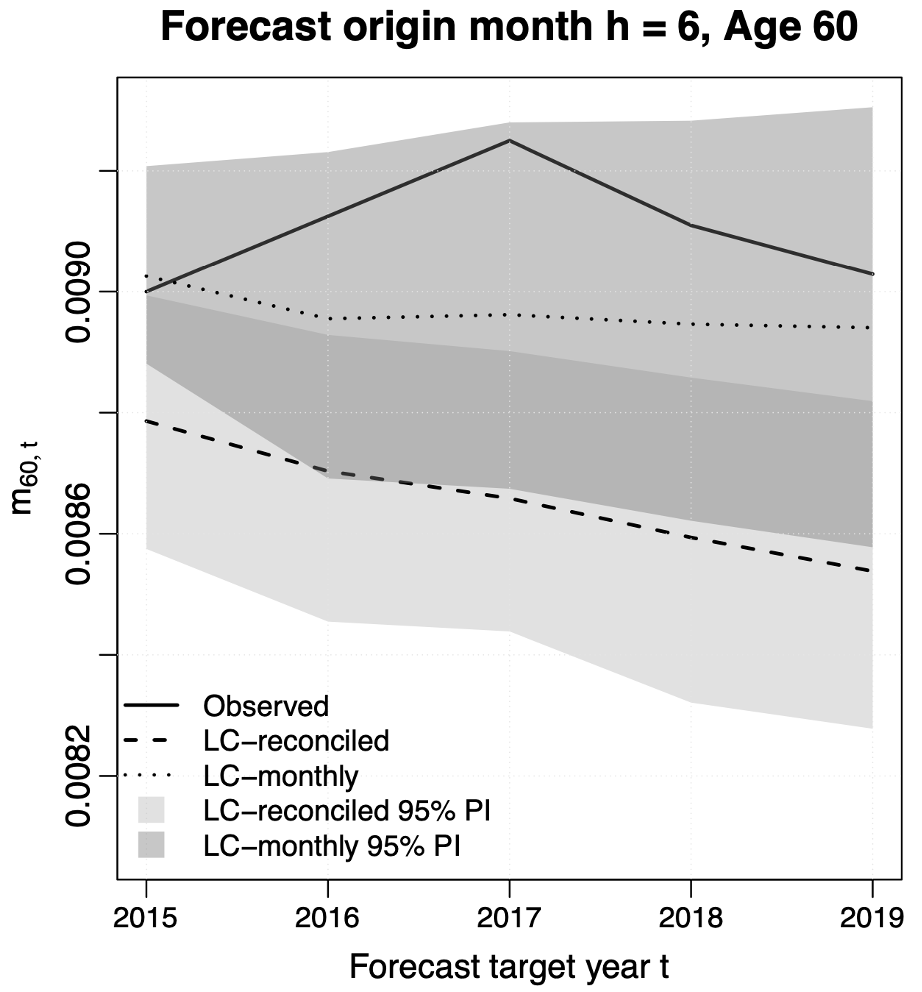}
\includegraphics[width=0.49\textwidth]{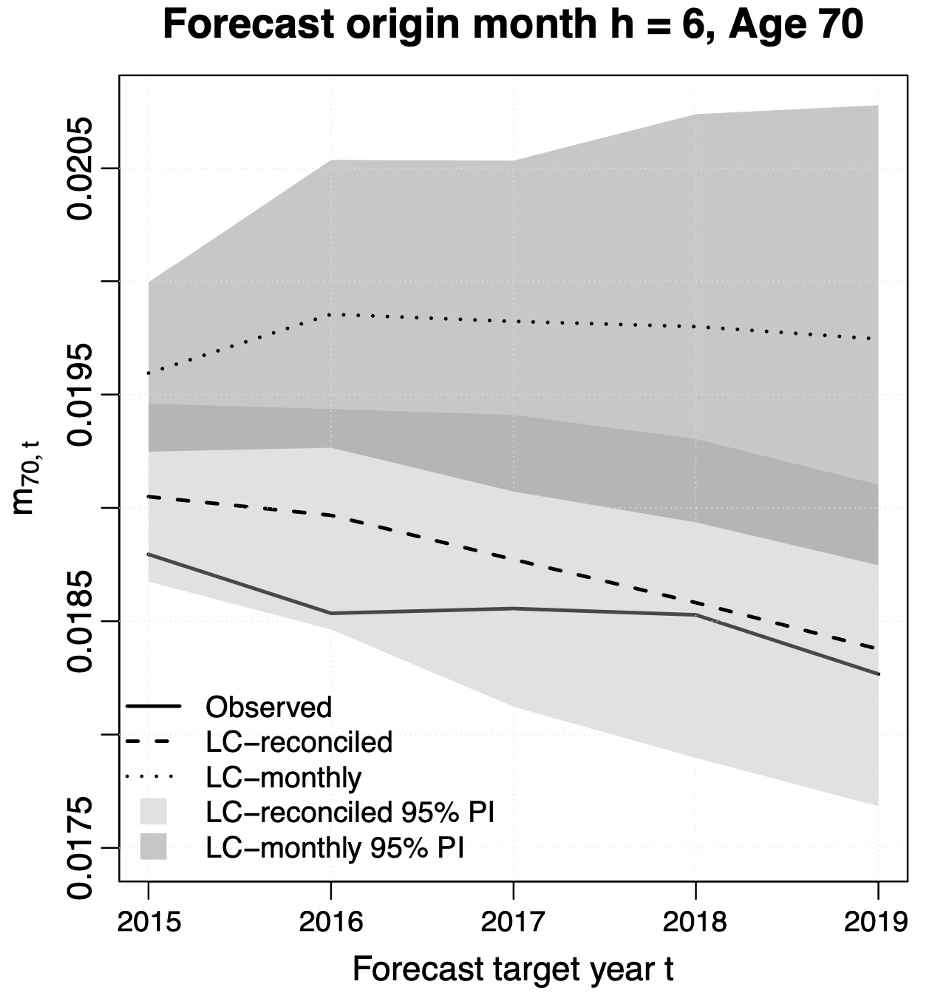}

\end{figure}

A further contributor is the reconciliation step itself. Temporal reconciliation produces forecasts by forming constrained weighted combinations of the annual and monthly base forecasts. This weighted average tends to smooth idiosyncratic variation that is inconsistent with the temporal aggregation constraint, thereby reducing dispersion in the reconciled forecast distribution. In addition, because the reconciliation weights are estimated from one-step-ahead errors and then reused across horizons, reconciled prediction intervals may be relatively tight at longer horizons if uncertainty growth is not fully reflected in the weight matrix. Figure~\ref{fig:prediction_interval_6ages_lc} illustrates this effect by comparing the 95\% prediction intervals from the LC--monthly and LC--reconciled forecasts. The reconciled intervals are much narrower across the two displayed ages. Consequently, narrower reconciled LC intervals should be interpreted as the result of the two-stage estimation and the reconciliation step, rather than as a guarantee of superior probabilistic calibration.

\section{Conclusions}
\label{sec:conclude}
This paper develops two practical strategies for combining monthly and annual mortality information for forecasting. The first is a MF-SS model in which a latent monthly period factor governs dynamics at both frequencies, allowing monthly and annual observations to inform the underlying common factor during estimation. The second strategy fits separate LC models to the annual series and to the scaled monthly series, and then applies an ex post temporal reconciliation step so that forecasts incorporate information from both monthly and annual levels. 

Both strategies deliver annual forecasts directly from an annual model or indirectly by aggregating forecasts generated at the monthly level. Empirically, annual forecasts constructed from monthly predictions are typically more accurate, except for nowcasts made very early in a calendar year. At the start of the year, the monthly approach has accumulated little intra-year information, whereas the annual model can provide a more stable signal.

We find that temporal reconciliation substantially improves the independently estimated LC forecasts, but has little effect on MF-SS forecasts. This pattern is consistent with the MF-SS framework already pooling information across frequencies through the shared latent factor. Even after reconciliation, the LC forecasts in our setting generally remain slightly less accurate than the MF-SS forecasts aggregated from the monthly forecasts. In addition, prediction intervals obtained from the reconciled LC approach can be overly narrow, suggesting that the two-stage procedure and the reconciliation may lead to understating forecast uncertainty.

Overall, our results favor the MF-SS model when both point accuracy and a more cautious assessment of uncertainty matter. If the primary objective is point forecasting and computational simplicity, however, the reconciled LC approach can provide a reasonable and less computationally intensive alternative.

In our empirical analysis, the annual and monthly series share the same estimation window. In practice, annual data often extend further back in time, while monthly data may be shorter and may also end earlier because of reporting delays or revisions. Both approaches can be implemented under such misalignment, but the mixed-frequency state-space model accommodates it naturally by treating unavailable observations as missing and continuing to update latent states with the information that is observed. A useful next step is to quantify the value of long annual histories for improving short-horizon and long-horizon forecasts when monthly data are limited.

The MF-SS framework also admits several extensions that could broaden its usefulness. One direction is to move beyond all-cause mortality and model cause-specific mortality jointly across causes, allowing for shared period dynamics and cause-specific deviations. Another direction is multi-country modeling, where both data coverage and reporting frequency differ across countries. A hierarchical or multi-level version of the model could pool information across countries through shared latent factors, improving inference on intra-year dynamics and short-term forecasts for countries with sparse or lower-frequency data, while still allowing country-specific patterns. 

Finally, reconciliation methods could be improved to better reflect dependence across temporal resolutions when combining independently generated forecasts. In particular, developing reconciliation procedures that incorporate cross-frequency error covariances and yield better-calibrated prediction intervals would make these approaches more reliable for risk assessment.

\bibliography{bib}

\appendix
\section{Likelihood and EM details}
\label{app:em_details}

\subsection{Time varying selection matrix}

In the main text, we described the mixed-frequency setting by explicitly partitioning the full observation vector into observed and unobserved components, $\bm{Y}_\tau=(\bm{Y}^{\mathrm{obs}}_\tau,\bm{Y}^{\mathrm{mis}}_\tau)$, to emphasise that the EM $Q$-function involves expectations with respect to the joint conditional distribution of the latent states and the unobserved entries of $\bm{Y}_{1:T}$. In this appendix, we use an equivalent but more compact notation based on a time-varying selection matrix. This formulation allows for cleaner mathematical expressions of the likelihoods. 

Let $\bm{M}_\tau$ be a time-varying selection matrix that extracts the observed components at month $\tau$, and define
\[
\bm{Y}^{\mathrm{obs}}_\tau=\bm{M}_\tau\bm{Y}_\tau,\qquad
\bm{A}_\tau=\bm{M}_\tau\bm{A},\qquad
\bm{B}_\tau=\bm{M}_\tau\bm{B},\qquad
\bm{R}_\tau=\bm{M}_\tau\bm{R}\bm{M}_\tau^\top,
\]
where $d_\tau=|\bm{Y}^{\mathrm{obs}}_\tau|$ may vary with $\tau$ (e.g., $d_\tau=n_a$ in intra-year months and $d_\tau=2n_a$ at year-end).

\subsection{Filtering and smoothing distributions in the E-step}
\label{app:kf_update}

In the prediction step, we obtain the conditional distribution of the latent state at time $\tau$ as
\[(\bm{k}_\tau \mid \bm{Y}^{\mathrm{obs}}_{1:\tau-1},\bm{\Theta}^{(i)})
  \;\sim\;
  \mathcal{N}\!\big(\bm{k}_{\tau|\tau-1},\,\bm{P}_{\tau|\tau-1}\big),
  \]
  \[
\bm{k}_{\tau|\tau-1}=\bm{H}\bm{k}_{\tau-1|\tau-1}+\bm{u},
\qquad
\bm{P}_{\tau|\tau-1}=\bm{H}\bm{P}_{\tau-1|\tau-1}\bm{H}^\top+\bm{Q},
\qquad
\bm{Q}=\sigma_w^2\bm{G}\bm{G}^\top.
\]

The observation innovation and its covariance are then
\[
\bm{v}_\tau=\bm{Y}^{\mathrm{obs}}_\tau-\bm{A}_\tau-\bm{B}_\tau \bm{k}_{\tau|\tau-1},
\qquad
\bm{S}_\tau=\bm{B}_\tau \bm{P}_{\tau|\tau-1}\bm{B}_\tau^\top+\bm{R}_\tau.
\]

In the update step, given the new observation, the conditional distribution of the latent state is expressed as 
 \[  (\bm{k}_\tau \mid \bm{Y}^{\mathrm{obs}}_{1:\tau},\bm{\Theta}^{(i)})
\sim
\mathcal{N}\!\big(\bm{k}_{\tau|\tau},\,\bm{P}_{\tau|\tau}\big),\]
\[
\bm{K}_\tau=\bm{P}_{\tau|\tau-1}\bm{B}_\tau^\top \bm{S}_\tau^{-1},
\qquad
\bm{k}_{\tau|\tau}=\bm{k}_{\tau|\tau-1}+\bm{K}_\tau \bm{v}_\tau,
\qquad
\bm{P}_{\tau|\tau}=\bm{P}_{\tau|\tau-1}-\bm{K}_\tau \bm{S}_\tau \bm{K}_\tau^\top.
\]
The smoother then provides $(\widetilde{\bm{k}}_\tau,\widetilde{\bm{P}}_\tau,\widetilde{\bm{P}}_{\tau,\tau-1})$ required in Appendix \ref{app:Q_func}; see \citet{koopman2012} for standard formulas.

\subsection{Observed-data log-likelihood}
\label{app:obs_ll}
The observed-data log-likelihood can be evaluated efficiently via the Kalman filter using the one-step-ahead forecast innovation representation.

The observed-data log-likelihood is
\begin{equation}
\label{eq:ll_innov_app}
\ell(\bm{\Theta})
=
\log f(\bm{Y}^{\mathrm{obs}}_{1:T}\mid \bm{\Theta})
=
-\tfrac{1}{2}\sum_{\tau=1}^{T}
\left[
d_\tau\log(2\pi)+\log|\bm{S}_\tau|+\bm{v}_\tau^\top\bm{S}_\tau^{-1}\bm{v}_\tau
\right].
\end{equation}
We monitor \eqref{eq:ll_innov_app} across EM iterations as an additional convergence diagnostic.

\subsection{Complete-data log-likelihood}
\label{app:complete_ll}
In our implementation, we define the complete data as the fixed-dimension observation sequence
$\bm{Y}_{1:T}=\{\bm{Y}_\tau\}_{\tau=1}^T$ together with the latent state sequence $\bm{k}_{0:T}$.
Let $\bm{\Theta}=\{\bm{A},\bm{B},\bm{H},\bm{R},\bm{u},\sigma_w^2\}$. Up to an additive constant, the complete-data log-likelihood decomposes into initial-state, observation, and state terms:
\begin{equation}
\label{eq:complete_ll_app}
\begin{aligned}
\log f(\bm{Y}_{1:T},\bm{k}_{0:T}\mid \bm{\Theta})
&=
-\tfrac{1}{2}\Big[
(\bm{k}_0-\bm{k}_{0|0})^\top \bm{P}_{0|0}^{-1}(\bm{k}_0-\bm{k}_{0|0})
+\log|\bm{P}_{0|0}|
\Big]\\
&\quad
-\tfrac{1}{2}\sum_{\tau=1}^{T}
\Big[
\bm{e}_\tau^\top \bm{R}^{-1}\bm{e}_\tau + \log|\bm{R}|
\Big]
-\tfrac{1}{2}\sum_{\tau=1}^{T}
\Big[
\tfrac{w_\tau^2}{\sigma_w^2}+\log\sigma_w^2
\Big],
\end{aligned}
\end{equation}
where the observation residual is
\[
\bm{e}_\tau = \bm{Y}_\tau-\bm{A}-\bm{B}\bm{k}_\tau,
\]
and the scalar state innovation is
\[
w_\tau=\bm{G}^\top\big(\bm{k}_\tau-\bm{H}\bm{k}_{\tau-1}-\bm{u}\big).
\]

\subsection{The $Q$-function and required smoothing moments}
\label{app:Q_func}

We treat the full sequence $\bm{Y}_{1:T}$ as part of the complete data. Accordingly, at EM iteration $i$ the $Q$-function is
\begin{equation}
\label{eq:Q_fullY_M}
Q(\bm{\Theta}\mid \bm{\Theta}^{(i)})
=
\mathbb{E}\!\left[
\log f(\bm{Y}_{1:T},\bm{k}_{0:T}\mid \bm{\Theta})
\ \Big|\ 
\{\bm{Y}^{\mathrm{obs}}_\tau\}_{\tau=1}^{T},\bm{\Theta}^{(i)}
\right],
\end{equation}
where the expectation is taken with respect to the joint conditional distribution of the latent states and the unobserved entries of $\bm{Y}_{1:T}$ implied by the model and the linear restrictions
$\bm{Y}^{\mathrm{obs}}_\tau=\bm{M}_\tau\bm{Y}_\tau$.

\paragraph{Smoothing moments of the latent states.}
Because the model is linear--Gaussian, the smoothing distribution
$p(\bm{k}_{0:T}\mid \{\bm{Y}^{\mathrm{obs}}_\tau\}_{\tau=1}^T,\bm{\Theta}^{(i)})$
is Gaussian and is characterized by the Kalman filter/smoother outputs:
\[
\widetilde{\bm{k}}_\tau=\mathbb{E}[\bm{k}_\tau\mid \bm{Y}^{\mathrm{obs}}_{1:T}],
\qquad
\widetilde{\bm{P}}_\tau=\mathrm{Var}(\bm{k}_\tau\mid \bm{Y}^{\mathrm{obs}}_{1:T}),
\qquad
\widetilde{\bm{P}}_{\tau,\tau-1}=\mathrm{Cov}(\bm{k}_\tau,\bm{k}_{\tau-1}\mid \bm{Y}^{\mathrm{obs}}_{1:T}).
\]

\paragraph{Conditional moments of the full observation vector $\bm{Y}_\tau$.}
The full measurement equation is written as
\[
\bm{Y}_\tau=\bm{A}+\bm{B}\bm{k}_\tau+\bm{\varepsilon}_\tau,\qquad
\bm{\varepsilon}_\tau\sim\mathcal{N}(\bm{0},\bm{R}).
\]
Conditional on $\bm{k}_\tau$ and the observed restriction $\bm{Y}^{\mathrm{obs}}_\tau=\bm{M}_\tau\bm{Y}_\tau$, the distribution of $\bm{Y}_\tau$ is Gaussian with mean and covariance
\begin{align}
\label{eq:Y_cond_mean}
\mathbb{E}[\bm{Y}_\tau \mid \bm{k}_\tau,\bm{Y}^{\mathrm{obs}}_\tau]
&=
\bm{\mu}_\tau(\bm{k}_\tau)
+
\bm{K}^{Y}_\tau\Big(\bm{Y}^{\mathrm{obs}}_\tau-\bm{M}_\tau\bm{\mu}_\tau(\bm{k}_\tau)\Big),\\
\label{eq:Y_cond_var}
\mathrm{Var}(\bm{Y}_\tau \mid \bm{k}_\tau,\bm{Y}^{\mathrm{obs}}_\tau)
&=
\bm{V}^{Y}_\tau
=
\bm{R}-\bm{R}\bm{M}_\tau^\top(\bm{M}_\tau\bm{R}\bm{M}_\tau^\top)^{-1}\bm{M}_\tau\bm{R},
\end{align}
where $\bm{\mu}_\tau(\bm{k}_\tau)=\bm{A}+\bm{B}\bm{k}_\tau$ and
\[
\bm{K}^{Y}_\tau=\bm{R}\bm{M}_\tau^\top(\bm{M}_\tau\bm{R}\bm{M}_\tau^\top)^{-1}.
\]
When all components of $\bm{Y}_\tau$ are observed, $\bm{M}_\tau=\bm{I}$ and $\bm{V}^{Y}_\tau=\bm{0}$, so $\bm{Y}_\tau$ is known exactly.

Integrating \eqref{eq:Y_cond_mean}--\eqref{eq:Y_cond_var} over the smoothing distribution of $\bm{k}_\tau$ yields the moments of $\bm{Y}_\tau$ required by the $Q$-function:
\begin{align}
\label{eq:Y_mean_smooth}
\overline{\bm{Y}}_\tau
:=
\mathbb{E}[\bm{Y}_\tau\mid \bm{Y}^{\mathrm{obs}}_{1:T}]
&=
\bm{A}+\bm{B}\widetilde{\bm{k}}_\tau
+\bm{K}^{Y}_\tau\Big(\bm{Y}^{\mathrm{obs}}_\tau-\bm{M}_\tau(\bm{A}+\bm{B}\widetilde{\bm{k}}_\tau)\Big),\\[4pt]
\label{eq:Y_var_smooth}
\mathrm{Var}(\bm{Y}_\tau\mid \bm{Y}^{\mathrm{obs}}_{1:T})
&=
\bm{V}^{Y}_\tau
+
(\bm{I}-\bm{K}^{Y}_\tau\bm{M}_\tau)\bm{B}\widetilde{\bm{P}}_\tau\bm{B}^\top(\bm{I}-\bm{K}^{Y}_\tau\bm{M}_\tau)^\top.
\end{align}
Consequently,
\[
\mathbb{E}[\bm{Y}_\tau\bm{Y}_\tau^\top\mid \bm{Y}^{\mathrm{obs}}_{1:T}]
=
\mathrm{Var}(\bm{Y}_\tau\mid \bm{Y}^{\mathrm{obs}}_{1:T})
+\overline{\bm{Y}}_\tau\,\overline{\bm{Y}}_\tau^\top.
\]
Moreover, the cross-moment needed in the quadratic expansion of the complete-data log-likelihood is
\begin{equation}
\label{eq:kY_cross}
\mathbb{E}[\bm{k}_\tau\bm{Y}_\tau^\top\mid \bm{Y}^{\mathrm{obs}}_{1:T}]
=
\widetilde{\bm{k}}_\tau\,\overline{\bm{Y}}_\tau^\top
+\widetilde{\bm{P}}_\tau\,\bm{B}^\top(\bm{I}-\bm{K}^{Y}_\tau\bm{M}_\tau)^\top.
\end{equation}

Together with the standard smoothing moments
$\mathbb{E}[\bm{k}_\tau\mid \bm{Y}^{\mathrm{obs}}_{1:T}]$, $\mathbb{E}[\bm{k}_\tau\bm{k}_\tau^\top\mid \bm{Y}^{\mathrm{obs}}_{1:T}]$ and
$\mathbb{E}[\bm{k}_\tau\bm{k}_{\tau-1}^\top\mid \bm{Y}^{\mathrm{obs}}_{1:T}]$,
the expressions \eqref{eq:Y_mean_smooth}--\eqref{eq:kY_cross} provide all expectations involving $\bm{Y}_\tau$ and $\bm{k}_\tau$ required to evaluate $Q(\bm{\Theta}\mid\bm{\Theta}^{(i)})$ and to derive closed-form M-step updates.

\subsection{Estimation procedure of the mixed-frequency state-space model}
\label{app:algorithm}
Algorithm \ref{alg:emkf} summarises the estimation procedure of the mixed-frequency state-space model. We set initial state moments to $\bm{k}_{0|0}=0$ and $\bm{P}_{0|0}=1$. The convergence threshold is set to $1\times10^{-6}$. The algorithm converges in approximately 25 iterations on average.

\begin{algorithm}[H]
\caption{Expectation--Maximization with Kalman Filter and Smoother}
\label{alg:emkf}
\SetAlgoLined
\DontPrintSemicolon

\textbf{Given:} observed series $\{\bm{Y}^{\mathrm{obs}}_\tau\}_{\tau=1}^{T}$; tolerance $c$; maximum iterations $I_{\max}$.\\
\textbf{Initialize:} choose $\bm{\Theta}^{(0)}=\{\bm{A}^{(0)},\bm{B}^{(0)},\bm{H}^{(0)},\bm{R}^{(0)},\bm{u}^{(0)},(\sigma_w^2)^{(0)}\}$ and initial state moments $(\bm{k}_{0|0},\bm{P}_{0|0})$, and set $i=0$ and $\Delta$ to a large number.\\

\While{$\Delta>c$ \textbf{and} $i<I_{\max}$}{

\BlankLine
\textbf{E-step (state estimation under missing observations).}\;

\begin{itemize}
    \item Run the Kalman filter sequentially, for $\tau=1,\ldots, T$, using $\bm{\Theta}^{(i)}$ and condition only on the observed information $\bm{Y}^{\mathrm{obs}}_\tau$ (i.e., omit the missing annual entries in intra-year months).\;
    \item Apply the Kalman smoother to compute the smoothing distribution
$p(\bm{k}_{1:T}\mid \bm{Y}^{\mathrm{obs}}_{1:T},\bm{\Theta}^{(i)})$
and obtain the smoothed state moments (means, variances, and lag-one covariances). 
\item 
Compute the current $Q$-function value
\[
Q\!\left(\bm{\Theta}^{(i)}\mid \bm{\Theta}^{(i)}\right)
=
\mathbb{E}\!\left[
\log f(\bm{Y}_{1:T},\bm{k}_{1:T}\mid \bm{\Theta}^{(i)})
\ \big|\ 
\bm{Y}^{\mathrm{obs}}_{1:T},\bm{\Theta}^{(i)}
\right],
\]
where the expectation is taken with respect to unobserved quantities
$(\bm{k}_{1:T},\bm{Y}^{\mathrm{mis}}_{1:T})\mid \bm{Y}^{\mathrm{obs}}_{1:T},\bm{\Theta}^{(i)}$.\;
\end{itemize}
\Indm

\BlankLine
\textbf{M-step (parameter updating).}\;
\Indp
\begin{itemize}
    \item Update $\bm{\Theta}$ by maximizing $Q(\bm{\Theta}\mid \bm{\Theta}^{(i)})$ to obtain
\[
\bm{\Theta}^{(i+1)}=\{\bm{A}^{(i+1)},\bm{B}^{(i+1)},\bm{H}^{(i+1)},\bm{R}^{(i+1)},\bm{u}^{(i+1)},\sigma_w^{(i+1)}\},
\]
using the closed-form EM updates (linear--Gaussian case) based on the smoothing distribution from the E-step. Missing annual components are handled through the same conditional expectations implied by
$(\bm{k}_{1:T},\bm{Y}^{\mathrm{mis}}_{1:T})\mid \bm{Y}^{\mathrm{obs}}_{1:T},\bm{\Theta}^{(i)}$.
\item 
Compute the updated $Q$-function value
\[
Q\!\left(\bm{\Theta}^{(i+1)}\mid \bm{\Theta}^{(i)}\right)
=
\mathbb{E}\!\left[
\log f(\bm{Y}_{1:T},\bm{k}_{1:T}\mid \bm{\Theta}^{(i+1)})
\ \big|\ 
\bm{Y}^{\mathrm{obs}}_{1:T},\bm{\Theta}^{(i)}
\right].
\]
\end{itemize}
\Indm

\BlankLine
\textbf{Convergence and likelihood monitoring.}\;
\Indp
\begin{itemize}
    \item Evaluate the stepwise relative change in the $Q$-function:
\[
\Delta
=
\frac{
\left|Q(\bm{\Theta}^{(i+1)}\mid \bm{\Theta}^{(i)})-Q(\bm{\Theta}^{(i)}\mid \bm{\Theta}^{(i)})\right|
}{
\left|Q(\bm{\Theta}^{(i)}\mid \bm{\Theta}^{(i)})\right|
}.
\]
\item 
Compute the observed-data log-likelihood
\[
\ell(\bm{\Theta}^{(i+1)})
=
\log f(\bm{Y}^{\mathrm{obs}}_{1:T}\mid \bm{\Theta}^{(i+1)}).
\]
\end{itemize}

\BlankLine
\textbf{Set $i\leftarrow i+1$.}\;

}
\end{algorithm}

\section{Constructing annual mortality rates from simulated monthly rates}
\label{sec:annualise}
To obtain annual mortality rates from scaled monthly death probability, we require values of monthly deaths and exposures. 
Given a simulated path for scaled monthly death rates for year $t$, $\{\tilde{z}_{x,(t,h)}\}_{h=1,\ldots,12}$,
we proceed sequentially for $h=1,2,\ldots,12$ to compute the simulated aggregated monthly rate:
\begin{enumerate}
    \item Compute simulated monthly deaths using the definition of the scaled monthly series:
    \[
    \tilde{D}_{x,(t,h)} = P_{x,t}\cdot \exp\!\big(\tilde{z}_{x,(t,h)}\big),
    \]
    where $P_{x,t}$ is the observed or simulated population size at the beginning of year $t$. 

    \item Determine simulated monthly population and exposure.
    Let $\tilde{P}_{x,(t,h)}$ denote the projected/observed population size at the beginning of month $h$ in year $t$, and let $\tilde{\Delta}_{x,(t,h)}$ denote the forecasted net international migration for age $x$ in month $h$ of year $t$. 
    We update the population month by month and approximate exposure by assuming deaths occur at mid-month\footnote{Our intra-year population and exposure construction applies demographic accounting ideas on a monthly grid. Specifically, our population update is a monthly analogue of the update logic used in \citet{CB2020} for the annual frequency. For exposures, we follow the Human Mortality Database convention \citep{HMDProtocol} based on uniform timing of deaths, but apply it at the monthly level.}:
    \begin{align*}
        \tilde{P}_{x,(t,h+1)} &= \tilde{P}_{x,(t,h)} + \tilde{\Delta}_{x,(t,h+1)} - \tilde{D}_{x,(t,h)},\\
        \tilde{E}_{x,(t,h)} &= \tfrac{1}{2}\Big(\tilde{P}_{x,(t,h+1)} + \tilde{P}_{x,(t,h)}\Big),
    \end{align*}
    with initialization $\tilde{P}_{x,(t,0)} = P_{x,t}$.
    Net migration is approximated using the information the previous year and assuming that immigration occurs uniformly throughout the year. In particular, we set
    \[
    \tilde{\Delta}_{x,(t,h)}=\Delta_{x,(t-1,h)}
    =\frac{1}{12}\Big(P_{x+1,t}-P_{x,t-1}+D_{x,t-1}\Big),
    \]
    where $P_{x+1,t}$, $P_{x,t-1}$ and $D_{x,t-1}$ are observed or, when forecasting beyond the sample, replaced by their corresponding forecasts.


    \item Compute the aggregated monthly forecast.
    We form the simulated annual mortality rates implied by the monthly path:
    \[
    \tilde{m}_{x,t}
    =
    \frac{\sum_{h=1}^{12} \tilde{D}_{x,(t,h)}}{\sum_{h=1}^{12} \tilde{E}_{x,(t,h)}}.
    \]
\end{enumerate}

If required, simulated future monthly mortality  can also be easily computed:
 \[
    \tilde{m}_{x,(t,h)}
    =
    \frac{\tilde{D}_{x,(t,h)}}{\tilde{E}_{x,(t,h)}}.
    \]
    
\end{document}